\documentclass[12pt]{iopart}
\usepackage{graphicx}
\usepackage{amsfonts}
\usepackage[T1]{fontenc}
\newcommand{\Z}{\mathbb{Z}}
\newcommand{\N}{\mathbb{N}}

\begin{document}

\title[Waveguides in a quantum perspective - Collin and Delattre]{Waveguides in a quantum perspective}

\author{Eddy Collin$^{1,*}$, Alexandre Delattre$^1$}

\address{$^1$Institut Néel - CNRS/UGA, 25 rue des Martyrs, 38042 Grenoble cedex 9, France}
\address{$^*$Author to whom correspondence should be addressed}
\ead{eddy.collin@neel.cnrs.fr}
\vspace{10pt}
\begin{indented}
\item[] \today
\end{indented}

\begin{abstract}

Solid state quantum devices, operated at dilution cryostat temperatures, are relying on microwave signals to both drive and read-out their quantum states. 
These signals are transmitted into the cryogenic environment, out of it towards detection devices, or even 
between quantum systems by well-designed {\it waveguides}, almost lossless when made of superconducting materials. 
Here we report on the quantum theory that describes the simplest Cartesian-type geometries: parallel plates, and rectangular tubes. 
The aim of the article is twofold: first on a technical and pragmatic level, we provide a full and compact quantum description of the different traveling wave families supported by these guides. Second, on an ontological level, we interpret the results and discuss the {\it nature} of the light fields corresponding to each mode family. 
The concept of potential difference is extended from transverse electric-magnetic (TEM) waves to {\it all} configurations, by means of a specific {\it gauge fixing}. The generalized 
flux $\hat{\varphi}$ introduced in the context of quantum electronics becomes here essential: 
it is the scalar field, conjugate of a charge $\hat{Q}$, that confines light within the electrodes, let them be real or {\it virtual}. The gap in the dispersion relations of non-TEM waves turns out to be linked either to a potential energy necessary for the {\it photon confinement}, or to a kinetic energy arising from a {\it photon mass}. 
We finally compute the field zero-point fluctuations in every configuration. The theory is predictive: the lowest transverse magnetic (TM) modes should have {\it smaller} quantum noise than the higher ones, which at large wavevectors recover a conventional value similar to TEM and transverse electric (TE) ones.
Such low-noise modes might be particularly useful for the 
routing of quantum information.

\end{abstract}

%
\vspace{2pc}

\section{Introduction}
\label{intro}

Mesoscopic devices, realized in the clean room and operated in commercial dilution cryostats, 
have reached such a level of control that their quantum properties are nowadays considered as a resource for emerging {\it quantum technologies}. These include in particular quantum computers, built around quantum bits made of Josephson Junctions \cite{qbitsupra}, or quantum dots prepared in semiconductors \cite{qubitsemi}.
But other particularly promising systems must be cited, as for instance electrons confined on helium \cite{elecHe},  magnons in magnetic materials \cite{YIG}, or micro-mechanical structures \cite{mech}.
Some of these are seen as potential solutions for key elements of a {\it quantum network}, such as light-to-microwave photon converters based on mechanical membranes operated near their ground state of motion \cite{photonconv}.

Without aiming at exhaustiveness here, we want to point out a common feature of these quantum systems: they rely on the transmission of signals in-and-out in the microwave range. This prolific research area has been named circuit quantum electro-dynamics (cQED) \cite{ClerckDevoret,valraff,blencowe,nori}. It shares many concepts with quantum optics, and relies on theories common to both fields, such as {\it input-output theory} and {\it open quantum systems theory} \cite{gardinerzoller}.
Beyond the technological revolution that these realizations aim at installing in everyday life, experiments also tackle fundamental questions of quantum mechanics: about a decade ago, vacuum fluctuations of a microwave electric circuit have been measured through the Lamb shift of a quantum bit \cite{lambWalraff}, and violations of a type of Bell's inequalities applying to a quantum bit have been reported \cite{leggettgargQuantro}. A remarkable recent experiment even demonstrated the entanglement between two {\it distant} quantum bits separated by a 30$~$m long superconducting waveguide \cite{loopholewalraff}, proving experimentally the fundamental non-locality describing quantum networks.

Waveguides are thus a key element of these setups. 
But their quantum description is performed at an already rather high level, relying on the {\it telegrapher equations} \cite{ClerckDevoret,valraff,blencowe,nori}.
Intrinsically, two conductors are required, with voltages and currents propagating sinusoidally along the lines. Their amplitudes are quantized, and become conjugate variables which can be measured by the experimentalist; in the second quantization language, they derive from bosonic creation/annihilation operators, which appear then in the Hamiltonian formalism that describes the coupling of the light field to any type of system.   
The work we present here is performed at a lower level, actually the lowest possible one: we start {\it directly from the fields} obtained from Maxwell's equations, limiting the discussion to the simplest transmission geometries in Cartesian coordinates. 
The method itself shall certainly be extended to
 other types of waveguides.

Why is such an approach useful? 
On a pragmatic level, it gives a complete description of all types of waves encountered in these geometries: not only the common transverse electric-magnetic (TEM) which is usually referred to, but also transverse electric (TE) and transverse magnetic (TM). 
For TEM waves, the notion of potential difference
is well defined; this is actually {\it not} the case for TE and TM \cite{pozar}. We demonstrate how to extend this concept, by introducing a specific {\it electromagnetic gauge}.
Pursuing on a fundamental level, we calculate all {\it constants of motion}: energy $H$, momentum $\vec{P}$ and angular momentum $\vec{J}$. 
These are interpreted physically through the introduction of the {\it generalized 
flux} $\hat{\varphi}$ \cite{devoret97} and its quantum conjugate, the charge $\hat{Q}$. The scalar field $\hat{\varphi}$ is thus responsible for light confinement, and lives on the boundaries {\it even in the absence of electrodes}: we shall call them virtual. 
The gap in the dispersion relations of non-TEM waves 
acquires a physical meaning: a {\it potential energy} cost for TM waves, and a {\it photon mass} for TE.
In the latter case, the Hamiltonian and the propagation equations are of the {\it Klein-Gordon} type.
Quantizing properly the fields leads to  well-known textbook expressions, but with a peculiarity: the calculated quantum fluctuations of TM waves tend to vanish at small wavevectors, while they recover a more standard expression at large wavevectors. This prediction could be tested in experiments, comparing quantitatively the quantum noise of different modes in the same waveguide. Besides, this property could find a specific use for quantum information transmission.

The article is structured as follows: we start in {\bf Section 2} by reminding basics of Maxwell's theory \cite{pozar,EMbook,cohen}, introducing our notations. We then give explicitly the field expressions for all configurations encountered. 
In {\bf Section 3}, the corresponding charges and currents living on the electrodes are calculated, together with the constants of motion $H$, $\vec{P}$ and $\vec{J}$ \cite{cohen}.
We express them all in terms of the generalized 
flux $\varphi$, 
first introduced in Ref. \cite{devoret97}. The potentials (scalar and vector) corresponding to the fields are presented in {\bf Section 4}, together with the concepts of {\it gauge invariance and gauge fixing} \cite{cohen,rovelli}. The generic relationship between $\varphi$ and the potentials is derived. 
The  {\bf Section 5} is devoted to the quantization of the fields \cite{ClerckDevoret,valraff,blencowe,nori,cohen,landau}, leading to $\varphi \rightarrow \hat{\varphi}$. From the initial quadratures, we introduce the bosonic operators and calculate the zero-point fluctuations of the electric and magnetic fields. 
A summary of our results and perspectives for future works are given in the Conclusion.

\section{Basics: Maxwell's equations}
\label{Max}

		\begin{figure}[t!]
		\centering
	\includegraphics[width=17cm]{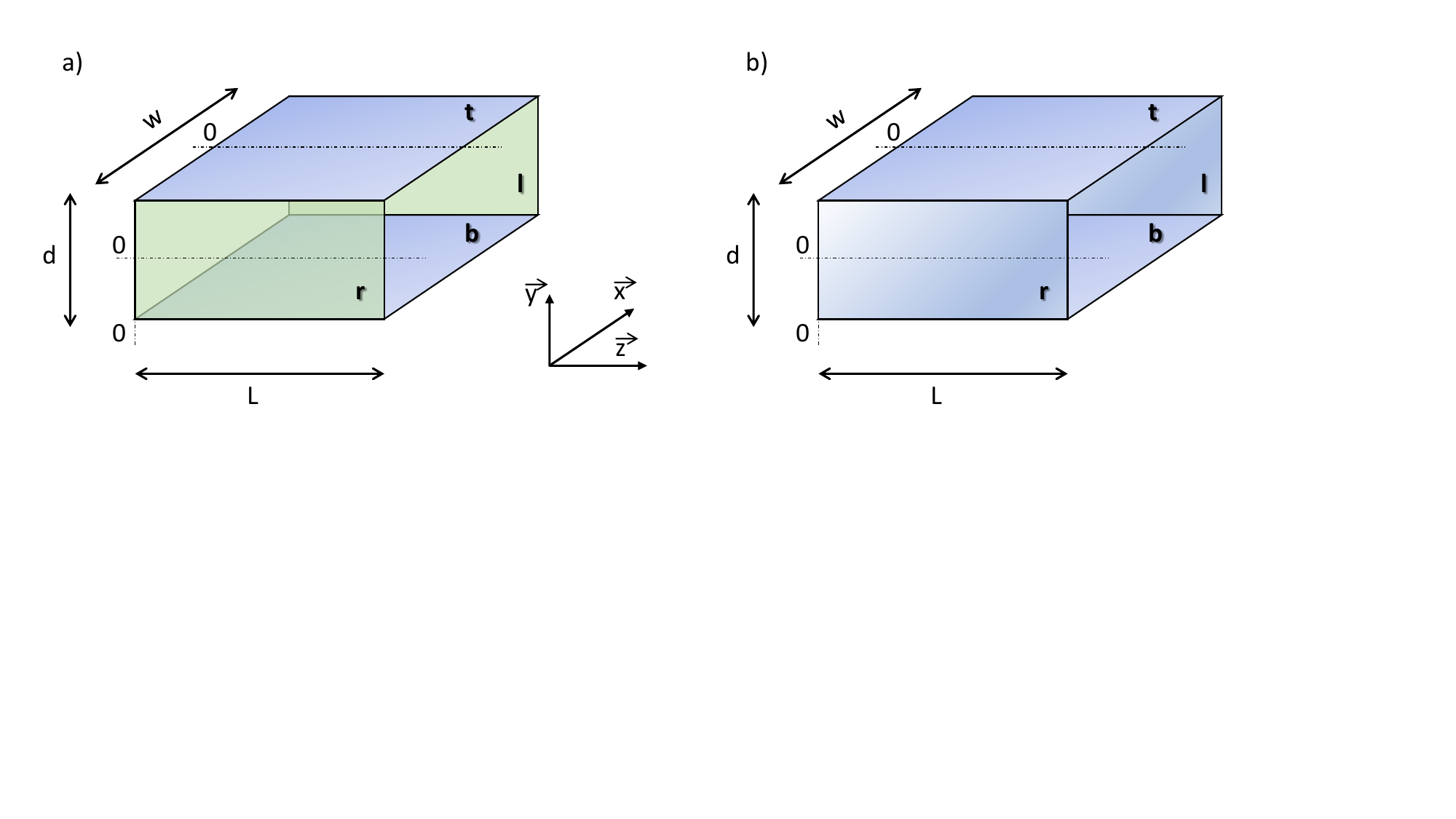}
    \vspace*{-5cm}			
			\caption{\small{ 
            a) Parallel plates transmission line. b) Rectangular transmission line. The electromagnetic field is enclosed within the colored surfaces: real electrodes in blue, and {\it virtual} ones in green. The width is $w$, height $d$ and length $L$. We name the electrodes {\bf t} (top), {\bf b} (bottom), {\bf l} (left) and {\bf r} (right). Positive direction of propagation: $\vec{z}$. }}
			\label{fig_1}
		\end{figure}

The geometries we are considering are shown in Fig. \ref{fig_1}. For the parallel plate case, we assume that $w \gg d$, such that fringing fields on the boundary (outside the green surfaces) can be ignored. For the rectangular guide, the dimensions can be any. But for the sake of concreteness, we can assume that the tube is oriented such that $w \geq d$, as on the Figure. Physical electrodes are depicted in blue; we call them "real", as opposed to the green boundary planes that we name "virtual" electrodes (for reasons that will appear below). We assume the inside of the guides to be empty space (no dielectric material). As such, the propagation of electromagnetic waves is described by the well-known Maxwell's equations \cite{pozar,EMbook,cohen}:
\begin{eqnarray}
\mbox{div} \, \vec{E}(\vec{r},t)&=&0 , \label{max1} \\
\mbox{div} \, \vec{B}(\vec{r},t)&=&0 , \\
\vec{\mbox{rot}} \, \vec{E}(\vec{r},t)&=& -\frac{\partial \vec{B}(\vec{r},t)}{ \partial t} , \\
\vec{\mbox{rot}} \, \vec{B}(\vec{r},t)&=& +\frac{1}{c^2} \frac{\partial \vec{E}(\vec{r},t)}{ \partial t} , \label{max4}
\end{eqnarray}
with $c$ the speed of light, and $\vec{E},\vec{B}$ (in Volts/meter and Tesla) the electric and magnetic fields respectively.
Note that since we aim at describing only quantum information transfer in waveguides (low-energy QED regime), these {\it linear} equations are all we need: non-linear corrections of the Euler-Heisenberg type due to {\it light-by-light} scattering can be safely ignored \cite{lightbylight}.  

A key ingredient here is {\it boundary conditions}. 
In their simplest form, they write for the real electrodes \cite{EMbook}:
\begin{eqnarray}
\vec{n} \times \vec{E}& = &0, \label{bound1} \\
\vec{n} \cdot \vec{E} &=& +\frac{\sigma_s}{\epsilon_0} , \\
\vec{n} \cdot \vec{B}& =& 0 , \\
\vec{n} \times \vec{B}&=& + \mu_0 \, \vec{j}_s , \label{bound4}
\end{eqnarray}
with $\vec{n}$ the vector normal to the electrode's surface, pointing towards the inside of the guide. 
$\epsilon_0$ is the vacuum permittivity, and $\mu_0$ the  permeability; one has $\epsilon_0 \mu_0 =1/c^2$. 
The $\times$ vectorial product therefore refers to the components of the fields parallel to the surface, while the scalar $\cdot$ refers to the perpendicular ones.  $\sigma_s$ and $\vec{j}_s$ correspond to the surface charge (in Coulomb/meter$^2$) and the surface current (in Ampere/meter) respectively.
What is assumed here for these expressions to be valid is that the thickness over which the charges/currents are confined is vanishingly small, and that the fields outside of the waveguide are zero. 
We also require losses in the electrodes to be negligible, which in practice would constrain the length $L$ we can consider without taking them into account. We shall not comment this point any further.
 
The former assumptions are easily satisfied in practice, for both good conductors (e.g. copper, silver, gold) and superconductors (e.g. aluminum, niobium operated at very low temperatures). 
The fields and charges/currents are confined on the inner surfaces within a lengthscale $\delta_s$ (skin depth $\propto 1/\sqrt{\omega}$, with $\omega$ angular frequency) for conductors, or $\lambda_L$ (London penetration depth) for superconductors, 
before they decay exponentially inside the material.
$\delta_s$ is typically of order $1~\mu$m around 5$~$GHz (at room temperature), while $\lambda_L$ ranges from 20$~$nm to 500$~$nm. 
In comparison, typical commercial waveguides operating in the microwave range have electrodes with thicknesses of order a (fraction of) millimeter, and dimensions $w,d$ of order a (fraction of) centimeter. 
Note indeed that remarkably, Eqs. (\ref{bound1}-\ref{bound4}) do not specify anything on the {\it nature} of the charges: these can be free electrons, or Cooper pairs.
In the quantum language, their statistics (bosons or fermions) is irrelevant here: the only thing that matters is that {\it they do couple to the electromagnetic field}.

We seek traveling wave solutions to this problem. As such, we introduce the following functions:
\begin{eqnarray}
f(z,t) & = & X \cos(\omega t - \beta z + \theta_0)+ Y \sin(\omega t - \beta z + \theta_0)  , \\
\tilde{f}(z,t) & = & X \sin(\omega t - \beta z + \theta_0)- Y \cos(\omega t - \beta z + \theta_0) ,
\end{eqnarray} 
with $\theta_0$ a phase that shall play no role, but just reminds that the chosen origin in space along $\vec{z}$ and in time $t$ is arbitrary.
$X$ and $Y$ are the two field quadratures, defined with no units.
These are the two measurable quantities that 
specify the electromagnetic field.
 $\beta$ is its wavevector, positive for a wave traveling in the direction of $\vec{z}$. 
We use periodic boundary conditions for our description, such that $\beta = 2 \pi \, l/L$, with $l \in \Z^*$ ($\beta \neq 0$, we exclude a non-propagating solution).
$\omega = c \, k$ is the angular frequency, with $k>0$ (we exclude a static solution). The {\it phase velocity} of the wave is thus $v_\phi = c \, k / | \beta | $ .
The electromagnetic field can then be written in a compact form, valid for all situations:
\begin{eqnarray}
E_x (\vec{r},t) & = & E_m \,\, g_{Ex} (x,y) f(z,t), \\
E_y (\vec{r},t) & = & E_m \,\, g_{Ey} (x,y) f(z,t), \\
E_z (\vec{r},t) & = & E_m \,\, g_{Ez} (x,y) \tilde{f}(z,t),
\end{eqnarray}
\vspace*{-7mm}
\begin{eqnarray}
B_x (\vec{r},t) & = & B_m \,\, g_{Bx} (x,y) f(z,t), \\
B_y (\vec{r},t) & = & B_m \,\, g_{By} (x,y) f(z,t), \\
B_z (\vec{r},t) & = & B_m \,\, g_{Bz} (x,y) \tilde{f}(z,t) .
\end{eqnarray}
$E_m$ and $B_m=E_m/c$ are two positive constants which carry the units of the fields (Volt/m and Tesla respectively), while the $g_i(x,y)$ normalized functions must be found in every specific case.
For given quadratures $X,Y$ the field magnitude is thus  encoded {\it only} in $E_m >0$, which must be defined by the theory.

\subsection{Parallel plates}
\label{plateG}

The simplest family of traveling waves is the transverse electric-magnetic (TEM) that exists within parallel plates \cite{pozar}.
The dispersion relation is linear, with simply $k = |\beta|$, and thus $v_\phi = c$.
 With the convention that the electric field on the top electrode is oriented {\it towards the inside} of the guide, the $g_i(x,y)$ functions are given in Tab. \ref{tab_1}.
In the following, we will systematically use normalizations matching this choice. 
Note that at the lateral boundaries (green planes in Fig. \ref{fig_1}), the electric field is parallel to the surface while the magnetic field is perpendicular. This is incompatible with the boundary conditions Eqs. (\ref{bound1}-\ref{bound4}), which means that in this case the "virtual" electrode is ill-defined: only the top and bottom electrodes must serve to describe the field confinement.

\begin{table}[h!]
\center
\caption{Modal functions for parallel plates TEM waves. }
\begin{tabular}{@{}lll}
\br
Function & & Expression \\
\mr
$g_{Ex} (x,y)$ & $=$ & $0$ \\
$g_{Ey} (x,y)$ & $=$ & $-1$ \\
$g_{Ez} (x,y)$ & $=$ & $0$ \\
$g_{Bx} (x,y)$ & $=$ & $+\mbox{sign}(\beta)$ \\
$g_{By} (x,y)$ & $=$ & $0$ \\
$g_{Bz} (x,y)$ & $=$ & $0$ \\
\br
\label{tab_1}
\end{tabular}
\end{table}
\normalsize

The second type of waves that are supported by parallel plates are transverse magnetic (TM) waves \cite{pozar}. 
In this case, the dispersion relation writes:
\begin{eqnarray}
k^2 & = & \beta^2 +k_c^2 , \\
k_c & = & \frac{n \pi}{d} ,
\end{eqnarray}
with $n \in \N^*$. There is a cutoff frequency $\omega_c = c \, k_c$ below which no TM wave can be excited. Each $n$ corresponds to a specific branch, which is denoted by TM$_n$. The modal functions are listed in Tab. \ref{tab_2}, taking again the convention that the electric field on the top electrode points towards the inside of the guide. As for TEM waves, the fields on the green planes are not compatible with electrodes' boundary conditions, and only the top and bottom real electrodes are to be considered.

\begin{table}[h!]
\center
\caption{Modal functions for parallel plates TM$_n$ waves. $n \neq 0$. }
\begin{tabular}{@{}lll}
\br
Function & & Expression \\
\mr
$g_{Ex} (x,y)$ & $=$ & $0$ \\
$g_{Ey} (x,y)$ & $=$ & $-(-1)^n \cos[k_c(y+d/2)] $ \\
$g_{Ez} (x,y)$ & $=$ & $+(-1)^n \frac{k_c}{\beta} \sin[k_c(y+d/2)]$ \\
$g_{Bx} (x,y)$ & $=$ & $+(-1)^n \frac{k}{\beta} \cos[k_c(y+d/2)]$ \\
$g_{By} (x,y)$ & $=$ & $0$ \\
$g_{Bz} (x,y)$ & $=$ & $0$ \\
\br
\label{tab_2}
\end{tabular}
\end{table}
\normalsize

The last family supported by parallel plates are transverse electric (TE) waves \cite{pozar}. As for TM waves, the dispersion relation is:
\begin{eqnarray}
k^2 & = & \beta^2 +k_c^2 , \\
k_c & = & \frac{n \pi}{d} ,
\end{eqnarray}
with $n \in \N^*$; we label again the branches TE$_n$. The same cutoff frequency $\omega_c = c \, k_c$ applies, below which no TE signal can propagate. The modal functions can be found in Tab. \ref{tab_3}.
These should be commented: the only nonzero component on the real electrodes is the {\it longitudinal magnetic field}, which is drastically different from the two preceding cases. Besides, on the lateral boundary surfaces, the electric field is perpendicular while the magnetic field is parallel: this is compatible with the boundary conditions Eqs. (\ref{bound1}-\ref{bound4}), which makes the virtual electrodes {\it meaningful}. 
It turns out that this field is {\it strictly equivalent} to the one obtained for the TE$_{n,m=0}$ wave of the rectangular guide (see Subsection \ref{rectG} below).
This justifies the terminology "virtual" electrode: the guide behaves essentially as if the electromagnetic field was confined in a closed pipe. We chose the convention such that the electric field on the left virtual electrode points towards the inside of the guide (with on top of it a $\sin[k_c(y+d/2)]$ modulation).
In order to be consistent with the next Subsection's definitions, we shall also make the formal substitution $E_m \rightarrow E_m'$ (the primes corresponding everywhere in the text to lateral electrodes, while the non-prime quantities refer to top/bottom).

\begin{table}[h!]
\center
\caption{Modal functions for parallel plates TE$_n$ waves. $n \neq 0$. }
\begin{tabular}{@{}lll}
\br
Function & & Expression \\
\mr
$g_{Ex} (x,y)$ & $=$ & $-\sin[k_c(y+d/2)]$ \\
$g_{Ey} (x,y)$ & $=$ & $0 $ \\
$g_{Ez} (x,y)$ & $=$ & $0 $ \\
$g_{Bx} (x,y)$ & $=$ & $0$ \\
$g_{By} (x,y)$ & $=$ & $- \frac{\beta}{k} \sin[k_c(y+d/2)]$ \\
$g_{Bz} (x,y)$ & $=$ & $- \frac{k_c}{k} \cos[k_c(y+d/2)]$ \\
\br
\label{tab_3}
\end{tabular}
\end{table}
\normalsize

Note finally that all the fields in Tabs. \ref{tab_1} - \ref{tab_3} are independent of $x$, which is a consequence of the assumption $w \gg d$.

\subsection{Rectangular guide}
\label{rectG}

The rectangular transmission line does not support TEM waves. It however supports TM$_{n,m}$ modes \cite{pozar}, with dispersion relation:
\begin{eqnarray}
k^2 & =& \beta^2 + k_{c}^2 , \\
k_c & = & \sqrt{k_{cx}^2 + k_{cy}^2} , \\
k_{cx} & = &  \frac{m \pi}{w}, \\
k_{cy} & = &  \frac{n \pi}{d}, 
\end{eqnarray}
and $m,n \in \N^*$. The cutoff frequency writes again $\omega_c= c \, k_c$, but depends now on the two indexes. The $g_i(x,y)$ functions are listed in Tab. \ref{tab_4}. The sign convention is chosen so as to match 
previous definitions, with an electric field on the top plate directed towards the inside of the guide (plus a $\sin[k_{cx}(x+w/2)]$ modulation).
But an equivalent choice can be made with the substitution $E_m \rightarrow E_m\,\!\!' = +(-1)^{n+m} E_m \, k_{cx}/k_{cy} $, leading to expressions given in Tab. \ref{tab_5}: in this case, the reference is taken on the left electrode, with an electric field pointing towards the inside of the guide (and presenting a $\sin[k_{cy}(y+d/2)]$ modulation).

\begin{table}[h!]
\center
\caption{Modal functions for rectangular guide TM$_{n,m}$ waves (referenced to top electrode). $n,m \neq 0$. }
\begin{tabular}{@{}lll}
\br
Function & & Expression \\
\mr
$g_{Ex} (x,y)$ & $=$ & $-(-1)^n \frac{k_{cx}}{k_{cy}}\cos[k_{cx}(x+w/2)]\sin[k_{cy}(y+d/2)] $ \\
$g_{Ey} (x,y)$ & $=$ & $-(-1)^n \sin[k_{cx}(x+w/2)]\cos[k_{cy}(y+d/2)] $ \\
$g_{Ez} (x,y)$ & $=$ & $+(-1)^n \frac{k_{cx}^2+k_{cy}^2}{k_{cy} \, \beta } \sin[k_{cx}(x+w/2)]\sin[k_{cy}(y+d/2)] $ \\
$g_{Bx} (x,y)$ & $=$ & $+(-1)^n \frac{k }{  \beta } \sin[k_{cx}(x+w/2)]\cos[k_{cy}(y+d/2)]$ \\
$g_{By} (x,y)$ & $=$ & $-(-1)^n \frac{k \, k_{cx}}{k_{cy} \,  \beta } \cos[k_{cx}(x+w/2)]\sin[k_{cy}(y+d/2)]$ \\
$g_{Bz} (x,y)$ & $=$ & $0$ \\
\br
\label{tab_4}
\end{tabular}
\end{table}
\normalsize

\begin{table}[h!]
\center
\caption{Modal functions for rectangular guide TM$_{n,m}$ waves (referenced to left electrode). $n,m \neq 0$. }
\begin{tabular}{@{}lll}
\br
Function & & Expression \\
\mr
$g_{Ex} (x,y)$ & $=$ & $-(-1)^m  \cos[k_{cx}(x+w/2)]\sin[k_{cy}(y+d/2)] $ \\
$g_{Ey} (x,y)$ & $=$ & $-(-1)^m \frac{k_{cy}}{k_{cx}} \sin[k_{cx}(x+w/2)]\cos[k_{cy}(y+d/2)] $ \\
$g_{Ez} (x,y)$ & $=$ & $+(-1)^m \frac{k_{cx}^2+k_{cy}^2}{k_{cx} \, \beta } \sin[k_{cx}(x+w/2)]\sin[k_{cy}(y+d/2)] $ \\
$g_{Bx} (x,y)$ & $=$ & $+(-1)^m \frac{k  \, k_{cy}}{  \, k_{cx} \,  \beta } \sin[k_{cx}(x+w/2)]\cos[k_{cy}(y+d/2)]$ \\
$g_{By} (x,y)$ & $=$ & $-(-1)^m \frac{k}{\beta } \cos[k_{cx}(x+w/2)]\sin[k_{cy}(y+d/2)]$ \\
$g_{Bz} (x,y)$ & $=$ & $0$ \\
\br
\label{tab_5}
\end{tabular}
\end{table}
\normalsize

The other mode family supported by the rectangular transmission line is TE$_{n,m}$ waves \cite{pozar}.
They verify again the 
dispersion relation:
\begin{eqnarray}
k^2 & =& \beta^2 + k_{c}^2 , \\
k_c & = & \sqrt{k_{cx}^2 + k_{cy}^2 } , \\
k_{cx} & = &  \frac{m \pi}{w}, \\
k_{cy} & = &  \frac{n \pi}{d}, 
\end{eqnarray}
with $m,n \in \N$ and at least one of them nonzero, and $\omega_c= c \, k_c$ (cutoff frequency).
Let us first consider the case $m=0$, and $n\neq0$. 
In this case, the obtained field is {\it exaclty the same} as the one of TE$_n$ waves confined within plates, see Tab. \ref{tab_3}, with simply $k_c = k_{cy}$. 
This is what justified our terminology of "virtual" electrodes for the green planes in Fig. \ref{fig_1}.
Choosing $n=0$, $m\neq0$ corresponds to an equivalent situation {\it rotated by 90$^\circ$ around the $\vec{z}$ axis}. The corresponding modal functions are given in Tab. \ref{tab_6}, with our electric field convention taken on the top (real) electrode; obviously $k_c = k_{cx}$.

\begin{table}[h!]
\center
\caption{Modal functions for rectangular guide TE$_{n=0,m}$ waves (referenced to top electrode). $m \neq 0$.}
\begin{tabular}{@{}lll}
\br
Function & & Expression \\
\mr
$g_{Ex} (x,y)$ & $=$ & $0$ \\
$g_{Ey} (x,y)$ & $=$ & $-\sin[k_{cx}(x+w/2)]$ \\
$g_{Ez} (x,y)$ & $=$ & $0$ \\
$g_{Bx} (x,y)$ & $=$ & $+\frac{\beta}{k} \sin[k_{cx}(x+w/2)]$ \\
$g_{By} (x,y)$ & $=$ & $0$ \\
$g_{Bz} (x,y)$ & $=$ & $+\frac{k_{cx}}{k} \cos[k_{cx}(x+w/2)]$ \\
\br
\label{tab_6}
\end{tabular}
\end{table}
\normalsize

We are left with TE$_{n,m}$ waves having both $n,m \neq 0$. 
The corresponding $g_i(x,y)$ functions are given in Tab. \ref{tab_7}. The convention is the same as above: the electric field on the top electrode is such that $\vec{E}$ on its surface points towards the inside of the guide, with a $\sin[k_{cx}(x+w/2)]$ modulation.
Equivalently, by means of the substitution $E_m \rightarrow E_m\,\!\!' = -(-1)^{n+m} E_m \, k_{cy}/k_{cx} $ we can rewrite these expressions referenced to the left electrode. This is given in Tab. \ref{tab_8}.

\begin{table}[h!]
\center
\caption{Modal functions for rectangular guide TE$_{n,m}$ waves (referenced to top electrode). $n,m \neq 0$. }
\begin{tabular}{@{}lll}
\br
Function & & Expression \\
\mr
$g_{Ex} (x,y)$ & $=$ & $+(-1)^n \frac{k_{cy}}{k_{cx}} \cos[k_{cx}(x+w/2)]\sin[k_{cy}(y+d/2)]$ \\
$g_{Ey} (x,y)$ & $=$ & $-(-1)^n \sin[k_{cx}(x+w/2)]\cos[k_{cy}(y+d/2)]$ \\
$g_{Ez} (x,y)$ & $=$ & $0$ \\
$g_{Bx} (x,y)$ & $=$ & $+(-1)^n \frac{\beta}{k } \sin[k_{cx}(x+w/2)]\cos[k_{cy}(y+d/2)]$ \\
$g_{By} (x,y)$ & $=$ & $+(-1)^n \frac{k_{cy} \, \beta}{k_{cx} \, k} \cos[k_{cx}(x+w/2)]\sin[k_{cy}(y+d/2)]$ \\
$g_{Bz} (x,y)$ & $=$ & $+(-1)^n \frac{k_{cx}^2+k_{cy}^2}{k_{cx} \, k} \cos[k_{cx}(x+w/2)]\cos[k_{cy}(y+d/2)]$ \\
\br
\label{tab_7}
\end{tabular}
\end{table}
\normalsize

The expressions given in this Section are all we  need to know in order to describe comprehensively the electromagnetic field within the waveguide. A given mode is thus characterized by a wavevector $\beta$ (or $l \in \Z^*$), and potentially branch indexes $m,n \in \N^*$. 

\begin{table}[h!]
\center
\caption{Modal functions for rectangular guide TE$_{n,m}$ waves (referenced to left electrode). $n,m \neq 0$. }
\begin{tabular}{@{}lll}
\br
Function & & Expression \\
\mr
$g_{Ex} (x,y)$ & $=$ & $-(-1)^m \cos[k_{cx}(x+w/2)]\sin[k_{cy}(y+d/2)]$ \\
$g_{Ey} (x,y)$ & $=$ & $+(-1)^m \frac{k_{cx}}{k_{cy}} \sin[k_{cx}(x+w/2)]\cos[k_{cy}(y+d/2)]$ \\
$g_{Ez} (x,y)$ & $=$ & $0$ \\
$g_{Bx} (x,y)$ & $=$ & $-(-1)^m \frac{k_{cx}\, \beta}{k_{cy} \, k} \sin[k_{cx}(x+w/2)]\cos[k_{cy}(y+d/2)]$ \\
$g_{By} (x,y)$ & $=$ & $-(-1)^m \frac{ \beta}{ k} \cos[k_{cx}(x+w/2)]\sin[k_{cy}(y+d/2)]$ \\
$g_{Bz} (x,y)$ & $=$ & $-(-1)^m \frac{k_{cx}^2+k_{cy}^2}{ k_{cy} \, k} \cos[k_{cx}(x+w/2)]\cos[k_{cy}(y+d/2)]$ \\
\br
\label{tab_8}
\end{tabular}
\end{table}
\normalsize

\section{Charges, currents, and constants of motion}
\label{charges}

Surface charges $\sigma_s$ and currents $\vec{j}_s$ are obtained through the boundary conditions Eqs. (\ref{bound1}-\ref{bound4}). These quantities obey {\it charge conservation} (a consequence of Maxwell's relations), and are therefore linked through the equation (with {\it div} the surface divergence):
\begin{equation}
\mbox{div} \, \vec{j}_s + \frac{\partial \sigma_s}{\partial t}=0 , \label{conserve}
\end{equation}
for each (real or virtual) electrode $s=t,b,l,r$ (standing for top, bottom, left and  right, respectively).

The constants of motion of the electromagnetic field (i.e. quantities independent of time $t$) are \cite{cohen}:
\begin{eqnarray}
H & = & \int_{z=0}^L \int_{x=-w/2}^{w/2}\int_{y=-d/2}^{d/2} \left[ \frac{1}{2} \epsilon_0 \vec{E}(\vec{r},t)^2 + \frac{1}{2} \frac{1}{\mu_0} \vec{B}(\vec{r},t)^2  \right] dx dy dz , \label{energy} \\
\vec{P} & = & \int_{z=0}^L \int_{x=-w/2}^{w/2}\int_{y=-d/2}^{d/2} \epsilon_0 \, \vec{E}(\vec{r},t) \times \vec{B}(\vec{r},t) \, dx dy dz , \label{momentum} \\
\vec{J} &=& \int_{z=0}^L \int_{x=-w/2}^{w/2}\int_{y=-d/2}^{d/2} \epsilon_0 \, \vec{r} \times \left[ \vec{E}(\vec{r},t) \times \vec{B}(\vec{r},t) \right] dx dy dz ,
\end{eqnarray}
with respectively $H$ the energy, $\vec{P}$ the momentum and $\vec{J}$ the angular momentum. For {\it all} the configurations studied here, we have $\vec{J}=0$.
In steady-state, the properties of the confined fields are fully characterized by these quantities; not mentioning the photon spin, or more properly {\it helicity} \cite{helicity}. Since in vacuum {\it two polarizations of light} (or degrees of freedom) are required to define this quantity, it seems obvious that here with only a single one the photon spin $\vec{S} $ {\it must be zero}. Concomitantly, writing $\vec{J} = \vec{L}+\vec{S}$, the orbital angular momentum of light $\vec{L}$ will also be zero; this might not be the case in other specific situations \cite{angularlight}. 
For these reasons, the angular momenta do not play any role here and will not be discussed any further.

At this stage, we shall introduce:
\begin{eqnarray}
C_d & = & \frac{\epsilon_0}{h_{ef\!f} } ,\\
C_d\,\!\!' & = & \frac{\epsilon_0}{h_{ef\!f}' } ,
\end{eqnarray}
the capacitance per unit surface for the top/bottom electrodes ($C_d$), and the one for the left/right electrodes ($C_d\,\!\!'$).
 Similarly, we have:
\begin{eqnarray}
L_d^{-1} & = & \frac{1}{\mu_0 \, h_{ef\!f} } ,\\
L_d\,\!\!'^{-1} & = & \frac{1}{\mu_0 \,h_{ef\!f}' } ,
\end{eqnarray}
the inverse inductance per unit surface, for top/bottom ($L_d$) and left/right ($L_d\,\!\!'$) electrodes.
The lengthscales $h_{ef\!f} \propto d$ and  $h_{ef\!f}' \propto w$ are derived below for each configuration.
We shall also introduce the generalized 
flux variables, which are defined {\it on} the electrodes. But for the time being, we will simply give the expressions based on dimensional analysis. The justification of this writing will appear in the following.
We thus have:
\begin{eqnarray}
\varphi(x,z,t) &=& \phi_m \,\, g_\phi (x) \tilde{f} (z,t) , \label{varphig1} \\
\varphi'(y,z,t) &=& \phi_m\,\!\!' \,\, g_{\phi\,'} (y) \tilde{f} (z,t) , \label{varphig2}
\end{eqnarray}
for the top/bottom ($\varphi$) and left/right ($\varphi '$) electrodes.
The two new modal functions $g_\phi (x)$ and $g_{\phi\,'} (y)$ will be given below as well for each configuration.
Similarly to the electromagnetic fields $\vec{E},\vec{B}$ themselves, the units of these expressions are contained in:
\begin{eqnarray}
\phi_m & = & \frac{E_m \, h_{ef\!f}}{\omega}, \\
\phi_m\,\!\!' & = & \frac{E_m\,\!\!' \, h_{ef\!f}' }{\omega}, 
\end{eqnarray} 
in Volt.seconds, directly proportional to the amplitude $E_m$. Note the $E_m\,\!\!'$ used in the second above mentioned equation, which ensures a proper normalization of modal functions when dealing with lateral electrodes.  
When these are virtual, they must obviously be meaningful for these expressions to apply (see previous Section).
In the present Section, we demonstrate that {\it all} the physical quantities linked to the light field can be derived from these generalized fluxes. 

\subsection{Parallel plates}

The TEM wave propagation is again the simplest situation, which brings $ h_{ef\!f}=d$ and $g_\phi (x)=1$; as expected, there is no $x$ dependence.
 From the boundary equations Eqs. (\ref{bound1}-\ref{bound4}), we obtain:
\begin{eqnarray}
\sigma_t & = & + C_d \frac{\partial \varphi(x,z,t)}{\partial t}, \\
\vec{j}_t & = &-  L_d^{-1} \frac{\partial \varphi(x,z,t)}{\partial z} \, \vec{z} ,
\end{eqnarray}
for the top electrode; the signs are reversed for the bottom one ($\sigma_b, \vec{j}_b$), 
which is just a consequence of the {\it symmetry} of the transverse electromagnetic field ($E_y$ and $B_x$ being actually constant in this case). Energy and momentum write, using Eqs. (\ref{energy},\ref{momentum}):
\begin{eqnarray}
H & = &\!\!\! \int_{z=0}^L \int_{x=-w/2}^{w/2}\left[ \frac{1}{2}\, C_d \left( \frac{\partial \varphi(x,z,t)}{\partial t}\right)^{\!\!2} + \frac{1}{2} \, L_d^{-1} \left( \frac{\partial \varphi(x,z,t)}{\partial z}\right)^{\!\!2} \right] dx dz , \label{enintTEM} \\
\vec{P} & = &\!\!\!  \int_{z=0}^L \int_{x=-w/2}^{w/2} \! C_d \left[-\frac{\partial \varphi(x,z,t)}{\partial t} \, \frac{\partial \varphi(x,z,t)}{\partial z} \right] dx dz \, \vec{z} .
\end{eqnarray}
The conservation law Eq. (\ref{conserve}) leads to:
\begin{equation}
\frac{\partial^2 \varphi(x,z,t)}{\partial z^2} - \frac{1}{c^2}  \frac{\partial^2 \varphi(x,z,t)}{\partial t^2} =0 ,
\end{equation}
which is nothing but the wave equation propagation for the flux variable $\varphi$, at the speed of light $c$. We see that the bracket in the energy integral above, Eq. (\ref{enintTEM}), is just the  surface charge/current energy density $H_d$:
\begin{equation}
H_d = \frac{1}{2}\, C_d^{-1} \,  \sigma_t^{ 2} + \frac{1}{2} \, L_d  \, \vec{j}_t^{ _, 2} ,
\end{equation}
which can be written equally well with the bottom electrode charge/current. For this configuration, the virtual electrode is not defined. \\

Consider now the TM$_n$ waves. In this case, we obtain $ h_{ef\!f}=d/2$ but still $g_\phi (x)=1$ (no $x$ dependence). Also, Eqs. (\ref{bound1}-\ref{bound4}) lead to:
\begin{eqnarray}
\sigma_t & = & + C_d \frac{\partial \varphi(x,z,t)}{\partial t}, \\
\vec{j}_t & = &-  L_d^{-1} \left(\frac{k}{\beta}\right)^{\!\!2} \frac{\partial \varphi(x,z,t)}{\partial z} \, \vec{z} ,
\end{eqnarray}
for the top electrode. The bottom electrode $\sigma_b,\vec{j}_b $ are obtained by multiplying the above by $-(-1)^n$: the top/bottom charges and currents are opposed for even $n$ (symmetric $E_y$, $B_x$ field profiles in $\vec{y}$ direction), but of same sign for $n$ odd (anti-symmetric). Energy and momentum are:
\begin{eqnarray}
H  & = &  \int_{z=0}^L \int_{x=-w/2}^{w/2} \left[ \frac{1}{2}\, C_d \left( \frac{\partial \varphi(x,z,t)}{\partial t}\right)^{\!\!2} + \frac{1}{2}\, C_d \left(\frac{k}{\beta}\right)^{\!\!2} \left(c\,k_c \right)^2 \varphi(x,z,t)^2 \right. \nonumber \\
& +& \left. \frac{1}{2} \, L_d^{-1}\left(\frac{k}{\beta}\right)^{\!\!4} \left( \frac{\partial \varphi(x,z,t)}{\partial z}\right)^{\!\!2} \right] dx dz , \label{enintTM} \\
\vec{P} &  =  &  \int_{z=0}^L \int_{x=-w/2}^{w/2} \! C_d \left(\frac{k}{\beta}\right)^{\!\!2} \left[-\frac{\partial \varphi(x,z,t)}{\partial t} \, \frac{\partial \varphi(x,z,t)}{\partial z} \right] dx dz \, \vec{z} .
\end{eqnarray}
The charge conservation leads also to a different propagation equation for the flux:
\begin{equation}
\frac{\partial^2 \varphi(x,z,t)}{\partial z^2} - \frac{1}{c^2} \left(\frac{\beta}{k}\right)^{\!\!2} \frac{\partial^2 \varphi(x,z,t)}{\partial t^2} =0 . \label{veloceTM}
\end{equation}
We see that all of these equations, compared to the TEM case, are renormalized by terms of the form $(v_\phi/c)^2 = (k/\beta)^2$  and $(v_\phi/c)^4 = (k/\beta)^4$. 
In particular, Eq. (\ref{veloceTM}) corresponds to a propagation {\it at a velocity} $v_\phi$ instead of $c$; the other quantities shall be commented later. 
The surface energy density $H_d$ appearing in Eq. (\ref{enintTM}) is also modified:
\begin{equation}
H_d = \frac{1}{2}\, C_d^{-1} \,  \sigma_t^{ 2} + \frac{1}{2} \, L_d  \, \vec{j}_t^{ _, 2} + \Delta(x,z,t) ,
\end{equation}
where we recognize the charge/current energy density, {\it plus a new term} $\Delta(x,z,t)$:
\begin{equation}
\Delta(x,z,t) = \frac{1}{2}\, C_d \left(\frac{k}{\beta}\right)^{\!\!2} \left(c\,k_c \right)^2 \varphi(x,z,t)^2. \label{equaDelta}
\end{equation}
We identify $\Delta(x,z,t)$ as being an {\it energy density linked to the cost} required for launching TM$_n$ waves into the guide.
The same expressions can be obtained from the bottom electrode charge and current; note also that the virtual electrodes are here ill-defined.\\

The last family of traveling waves present within parallel plates is TE$_n$ modes. For these, $ h_{ef\!f}'=w $ 
and the description is performed using the lateral (virtual) electrodes. We have $g_{\phi'} (y)=\sin[k_{c}(y+d/2)]$, and the boundary conditions can be used to define virtual charges and currents:
\begin{eqnarray}
\sigma_l & = & + C_d\,\!\!' \, \frac{\partial \varphi'(y,z,t)}{\partial t}, \label{sigl} \\
\vec{j}_l & = &-  L_d\,\!\!'^{-1} \, \frac{\partial \varphi'(y,z,t)}{\partial z} \, \vec{z}-  L_d\,\!\!'^{-1} \, \frac{\partial \varphi'(y,z,t)}{\partial y} \, \vec{y} , \label{jil}
\end{eqnarray}
for the left electrode; for the right one, invert all signs %
(the transverse field components $E_x$, $B_y$ being symmetric along $\vec{x}$, since independent of $x$).
These expressions resemble the TEM ones, but with a new peculiarity: there is a {\it transverse} current along $\vec{y}$.  
For the sake of completeness, let us look at the top/bottom (real) electrodes. We have:
\begin{eqnarray}
\sigma_b & = & 0, \\
\vec{j}_b & = &-  L_d^{-1} \,k_c \, \varphi(x,z,t) \, \vec{x} .
\end{eqnarray}
For the top electrode, $\sigma_t=0$ as well and $\vec{j}_t$ is obtained from $\vec{j}_b$ by a sign change $-(-1)^n$.
This writing requires to pose $ h_{ef\!f}=d $ and $g_{\phi} (x)=1$. 
What this tells us is that the real electrodes only host a transverse current. One can show that, at the plates extremities $x=\pm w/2$, $\vec{j}_b$ and $\vec{j}_t$ are continuous with $\vec{j}_l$ and $\vec{j}_r$: this mimics a {\it peripheral} current circulating around the electromagnetic field, even though two electrodes are virtual. This is reminiscent of the rectangular guide, which TE$_{n,m=0}$ family is strictly equivalent to the TE$_n$ one (see thereafter).
Energy and momentum write:
\begin{eqnarray}
H  & = &  \int_{z=0}^L \int_{y=-d/2}^{d/2} \left[ \frac{1}{2}\, C_d\,\!\!' \left( \frac{\partial \varphi'(y,z,t)}{\partial t}\right)^{\!\!2}  \right. \nonumber \\
& +& \left. \frac{1}{2} \, L_d\,\!\!'^{-1}  \left( \frac{\partial \varphi'(y,z,t)}{\partial z}\right)^{\!\!2} + \frac{1}{2}\,  L_d\,\!\!'^{-1}  \left( \frac{\partial \varphi'(y,z,t)}{\partial y}\right)^{\!\!2} \right] dy dz , \label{enintTE} \\
\vec{P} &  =  &  \int_{z=0}^L \int_{y=-d/2}^{d/2} \! C_d\,\!\!'  \left[-\frac{\partial \varphi'(y,z,t)}{\partial t} \, \frac{\partial \varphi'(y,z,t)}{\partial z} \right] dy dz \, \vec{z} ,
\end{eqnarray}
while the conservation equation Eq. (\ref{conserve}) brings:
\begin{equation}
\frac{\partial^2 \varphi'(y,z,t)}{\partial z^2} - \frac{1}{c^2}  \frac{\partial^2 \varphi'(y,z,t)}{\partial t^2} = +k_c^2 \varphi'(y,z,t) . \label{veloceTE}
\end{equation}
We see that in Eq. (\ref{veloceTE}), the transverse current is responsible for a new term $\propto \varphi'(y,z,t)$: this is exactly the Klein-Gordon propagation equation, and here $k_c \, \hbar /c = m  $ plays the role of a {\it mass}; this will be discussed in Section \ref{quantize}. 
The energy density in Eq. (\ref{enintTE}) is:
\begin{equation}
H_d' = \frac{1}{2}\, C_d\,\!\!'^{-1} \,  \sigma_l^{ 2} + \frac{1}{2} \, L_d\,\!\!'  \, \vec{j}_l^{\, 2} , \label{HdTE}
\end{equation}
which is nothing but the energy density of the virtual charges/currents. The extra term appearing in Eq. (\ref{enintTE}), as compared to the TEM case, is nothing but the energy density of the transverse current.
The same expressions can be written in terms of the right virtual charge/current. But note that the real electrodes, which are obviously necessary to ensure that the confined wave exists, {\it cannot be used} to describe it: the virtual electrodes are here essential. \\

But what does really happen in the corners of Fig. \ref{fig_1} a), where the blue real electrodes meet the green virtual planes?
There should be charge accumulation on one boundary, while there should be a deficit on the other (bouncing periodically at frequency $\omega$). These border defects should typically extend on a lengthscale $\sim d \ll w$, which is {\it not} captured in the modeling; one would require a more sophisticated analysis (like finite element simulation) in order to visualize it.
This irregularity in the microwave confining charge/current densities is also what is responsible for the energy leakage outside of the guide, through {\it fringing fields}. Our modeling assumes an ideal confinement of light, which is actually equivalent to having proper metallic walls in place of the virtual electrodes.

Pushing a bit further the analysis, one can wonder what the notion of "virtual electrode" really means.
Let us make a {\it gedankenexperiment}, by placing a non-invasive probe within the green planes of Fig. \ref{fig_1}. 
And let us for instance assume that this is a tiny metallic surface, oriented parallel to the green boundary, and in which we can measure small charges/currents. 
Since the $\vec{E},\vec{B}$ fields are consistent with metallic boundary conditions, {\it we should precisely measure} what is given by Eqs. (\ref{sigl},\ref{jil}).
The wording "virtual" is now clear: we are considering 
 charges and currents that we would have in an electrode, if there would be one. This is a fairly generic statement that can be applied to other electromagnetic problems, and which actually demonstrates how profound and important the boundary conditions Eqs. (\ref{bound1}-\ref{bound4}) are: 
they do not require anything about the nature of the charges at the boundary, as already pointed out, but also they do not even require that there is a {\it physical} electrode at this position in space.

\subsection{Rectangular guide}

TM$_{n,m}$ waves share the specific properties of the preceding TM$_n$ ones, as will become clear below.  
But here the lateral boundaries {\it physically exist}, and as such must be taken into account.
We have $ h_{ef\!f}=d/2 $, $g_{\phi} (x)=\sin[k_{cx}(x+w/2)]$ and $ h_{ef\!f}'=w/2 $, $g_{\phi'} (y)=\sin[k_{cy}(y+d/2)]$. The top/bottom boundary conditions lead to:
\begin{eqnarray}
\sigma_t & = & + C_d \frac{\partial \varphi(x,z,t)}{\partial t}, \\
\vec{j}_t & = &-  L_d^{-1} \left(\frac{k}{\beta}\right)^{\!\!2} \frac{\partial \varphi(x,z,t)}{\partial z} \, \vec{z} ,
\end{eqnarray}
and $\sigma_b$, $\vec{j}_b$ are obtained by multiplying both equations by $-(-1)^n$. Equivalent expressions are obtained 
for the left/right electrodes $\sigma_{l,r}$, $\vec{j}_{l,r}$  using primed quantities. This time, the symmetry between the facing electrodes is given by a factor $-(-1)^m$.
As for the parallel plate case, these signs are a simple consequence of the symmetry/anti-symmetry of the  transverse $\vec{E},\vec{B}$ fields in the direction normal to the electrodes.
Splitting adequately the integrals, we obtain for energy and momentum:
\begin{eqnarray}
\!\!\!\!\!\!\!\!\!\!\!\!\!\!\!\!\!\!\!\!\!\!\!\!\!\!\!\!\!  H  &\!\!\!\!\!\!\!\!\!\!\!\!\!\!\!\!\!\!\!\!\!\! = &\!\!\!\! \!\!\!\!\!\!\!\!\!\!\!\!  \int_{z=0}^L \int_{x=-w/2}^{w/2} \left[ \frac{1}{2}\, C_d \left( \frac{\partial \varphi(x,z,t)}{\partial t}\right)^{\!\!2} +  \frac{1}{4}\, C_d \left(1+\frac{k_{cx}^2}{k_{cy}^2}\right) \left(\frac{k}{\beta}\right)^{\!\!2} \left(c\,k_c \right)^2 \varphi(x,z,t)^2 \right. \nonumber \\
&\!\!\!\!\!\!\!\!\!\!\!\!\!\!\!\!\!\!\!\!\!\! +&\!\!\!\!\!\!  \!\!\!\!\!\!\!\!\!\! \left. \frac{1}{2} \, L_d^{-1}\left(\frac{k}{\beta}\right)^{\!\!4} \left( \frac{\partial \varphi(x,z,t)}{\partial z}\right)^{\!\!2} \right] dx dz \nonumber \\
&\!\!\!\!\!\!\!\!\!\!\!\!\!\!\!\!\!\!\!\!\!\! + &\!\!\!\!\! \!\!\!\!\!\!\!\!\!\!\! \int_{z=0}^L \int_{y=-d/2}^{d/2} \left[ \frac{1}{2}\, C_d\,\!\!' \left( \frac{\partial \varphi'(y,z,t)}{\partial t}\right)^{\!\!2} +  \frac{1}{4}\, C_d\,\!\!' \left(1+\frac{k_{cy}^2}{k_{cx}^2}\right) \left(\frac{k}{\beta}\right)^{\!\!2} \left(c\,k_c \right)^2 \varphi'(y,z,t)^2 \right. \nonumber \\
&\!\!\!\!\!\!\!\!\!\!\!\!\!\!\!\!\!\!\!\!\!\! +&\!\!\!\!\!\!\!\!\!\!\!\!\!\!\!\! \left. \frac{1}{2} \, L_d\,\!\!'^{-1}\left(\frac{k}{\beta}\right)^{\!\!4} \left( \frac{\partial \varphi'(y,z,t)}{\partial z}\right)^{\!\!2} \right] dy dz, \label{enintTM2} \\
\!\!\!\!\!\!\!\!\!\!\!\!\!\!\!\!\!\!\!\!\!\!\!\!\!\!\!\!\! \vec{P} &\!\!\!\!\!\!\!\!\!\!\!\!\!\!\!\!\!\!\!\!\!\!  =  &  \!\!\!\!\!\!\!\!\!\!\!\!\!\!\!\!  \int_{z=0}^L \int_{x=-w/2}^{w/2} \! C_d \left(\frac{k}{\beta}\right)^{\!\!2} \left[-\frac{\partial \varphi(x,z,t)}{\partial t} \, \frac{\partial \varphi(x,z,t)}{\partial z} \right] dx dz \, \vec{z} \nonumber \\
&   &\!\!\!\!\!\!\!\!\!\!\!\!\!\!\!\!\!\!\!\!\!\! +  \int_{z=0}^L \int_{y=-d/2}^{d/2} \! C_d\,\!\!' \left(\frac{k}{\beta}\right)^{\!\!2} \left[-\frac{\partial \varphi'(y,z,t)}{\partial t} \, \frac{\partial \varphi'(y,z,t)}{\partial z} \right] dy dz \, \vec{z} ,
\end{eqnarray}
which clearly presents the contribution from each pair, top/bottom (integrals on $x$ and $z$) and left/right (integrals on $y$ and $z$).
The propagation equation for the flux $\varphi$ is again obtained from charge conservation:
\begin{equation}
\frac{\partial^2 \varphi(x,z,t)}{\partial z^2} - \frac{1}{c^2} \left(\frac{\beta}{k}\right)^{\!\!2} \frac{\partial^2 \varphi(x,z,t)}{\partial t^2} =0 , \label{veloceTM2}
\end{equation}
and we recognize the $1/v_\phi^2$ velocity term. The same result stands for the $\varphi'$ flux. As for TM$_n$ waves, the energy densities $H_d, H_d'$ on each pair of electrodes correspond to the charge/current energy densities plus an extra term:
\begin{eqnarray}
\Delta(x,z,t) & = & \frac{1}{4}\, C_d \left(1+\frac{k_{cx}^2}{k_{cy}^2}\right) \left(\frac{k}{\beta}\right)^{\!\!2} \left(c\,k_c \right)^2 \varphi(x,z,t)^2 , \label{gap1} \\
\Delta'(y,z,t) & = & \frac{1}{4}\, C_d\,\!\!' \left(1+\frac{k_{cy}^2}{k_{cx}^2}\right) \left(\frac{k}{\beta}\right)^{\!\!2} \left(c\,k_c \right)^2 \varphi'(y,z,t)^2 ,\label{gap2}
\end{eqnarray}
with $\Delta'$ standing for the left/right pair. Note that even though we have split the writing between top/bottom and left/right electrodes, there is only {\it one} degree of freedom in the problem (described by the $X,Y$ quadratures). Actually, the surface integrals of the two energy addenda $\Delta, \Delta'$ are equal, which is why a factor $1/4$ appears in Eqs. (\ref{enintTM2},\ref{gap1},\ref{gap2}); note the difference with Eq. (\ref{equaDelta}). \\

The TE$_{n,m=0}$ waves are strictly equivalent to the TE$_n$ ones, treated above in the case of a parallel plate guide. The TE$_{n=0,m}$ waves are nothing but the same ones, rotated by 90$^\circ$ around the $\vec{z}$ axis. All the formulas are thus equivalent, using top/bottom electrodes (primed) quantities for the description.
The only difference being here that all electrodes are real ones, they are no virtual boundaries.
 We shall therefore discuss below only the TE$_{n,m}$ waves with $n,m \in \N^*$.
Their main specificities are the same as TE$_n$ ones.
As for transverse magnetic modes, we pose $ h_{ef\!f}=d/2 $, $g_{\phi} (x)=\sin[k_{cx}(x+w/2)]$ and $ h_{ef\!f}'=w/2 $, $g_{\phi'} (y)=\sin[k_{cy}(y+d/2)]$. The boundary conditions Eqs. (\ref{bound1}-\ref{bound4}) impose the currents and charges on top/bottom electrodes:
\begin{eqnarray}
\sigma_t & = & + C_d \, \frac{\partial \varphi(x,z,t)}{\partial t},  \\
\vec{j}_t & = &-  L_d^{-1} \, \frac{\partial \varphi(x,z,t)}{\partial z} \, \vec{z}-  L_d^{-1} \left(1+\frac{k_{cy}^2}{k_{cx}^2} \right) \, \frac{\partial \varphi(x,z,t)}{\partial x} \, \vec{x} , \label{jil2}
\end{eqnarray}
with a sign change $-(-1)^n$ for $\sigma_b, \vec{j}_b$.
For the left/right pair, the writing is:
\begin{eqnarray}
\sigma_l & = & + C_d\,\!\!' \, \frac{\partial \varphi'(y,z,t)}{\partial t},   \\
\vec{j}_l & = &-  L_d\,\!\!'^{-1} \, \frac{\partial \varphi'(y,z,t)}{\partial z} \, \vec{z}-  L_d\,\!\!'^{-1} \left(1+\frac{k_{cx}^2}{k_{cy}^2} \right) \, \frac{\partial \varphi'(y,z,t)}{\partial y} \, \vec{y} \label{jil3},  
\end{eqnarray}
with a sign change $-(-1)^m$ for the right charge/current densities.
As previously, the sign change in charges/currents of facing electrodes is linked to the symmetry/anti-symmetry of the transverse electromagnetic field profile.
 These expressions are similar to Eqs. (\ref{sigl},\ref{jil}), and the transverse components in Eqs. (\ref{jil2},\ref{jil3}) correspond to a peripheral current running around the guide. 
Judiciously splitting the volumic integral on the fields, we obtain for energy and momentum:
\begin{eqnarray}
\!\!\!\!\!\!\!\!\!\!\!\!\!\!\!\!\!  H  &\!\!\!\!\!\!\!\!\!\!\!\! = &\!\!\!\!\!\!\!  \int_{z=0}^L \int_{x=-w/2}^{w/2} \left[ \frac{1}{2}\, C_d \left( \frac{\partial \varphi(x,z,t)}{\partial t}\right)^{\!\!2}  \right. \nonumber \\
&\!\!\!\!\!\!\!\!\!\!\!\! +&\!\!\!\!\!\!\!  \left. \frac{1}{2} \, L_d^{-1} \left( \frac{\partial \varphi(x,z,t)}{\partial z}\right)^{\!\!2} +   \frac{1}{4} \, L_d^{-1} \left(1+ \frac{k_{cy}^2}{k_{cx}^2} \right)^{\!\!2} \left( \frac{\partial \varphi(x,z,t)}{\partial x}\right)^{\!\!2} \right] dx dz \nonumber \\
&\!\!\!\!\!\!\!\!\!\!\!\! + &\!\!\!\!\!\!\!  \int_{z=0}^L \int_{y=-d/2}^{d/2} \left[ \frac{1}{2}\, C_d\,\!\!' \left( \frac{\partial \varphi'(y,z,t)}{\partial t}\right)^{\!\!2}  \right. \nonumber \\
&\!\!\!\!\!\!\!\!\!\!\!\! +&\!\!\!\!\!\!\!  \left. \frac{1}{2} \, L_d\,\!\!'^{-1} \left( \frac{\partial \varphi'(y,z,t)}{\partial z}\right)^{\!\!2} +   \frac{1}{4} \, L_d\,\!\!'^{-1} \left( 1+ \frac{k_{cx}^2}{k_{cy}^2} \right)^{\!\!2} \left( \frac{\partial \varphi'(y,z,t)}{\partial y}\right)^{\!\!2}\right] dy dz, \label{enintTE2} \\
\!\!\!\!\!\!\!\!\!\!\!\!\!\!\!\!\! \vec{P} &\!\!\!\!\!\!\!\!\!\!\!\!  =  &\!\!\!\!\!\!\!    \int_{z=0}^L \int_{x=-w/2}^{w/2} \! C_d   \left[-\frac{\partial \varphi(x,z,t)}{\partial t} \, \frac{\partial \varphi(x,z,t)}{\partial z} \right] dx dz \, \vec{z} \nonumber \\
&   &\!\!\!\!\!\!\!\!\!\!\!   +  \int_{z=0}^L \int_{y=-d/2}^{d/2} \! C_d\,\!\!'   \left[-\frac{\partial \varphi'(y,z,t)}{\partial t} \, \frac{\partial \varphi'(y,z,t)}{\partial z} \right] dy dz \, \vec{z} .
\end{eqnarray}
The two sets of integrals in these equations correspond to the two pairs of electrodes. 
Charge conservation brings the propagation equation:
\begin{equation}
\frac{\partial^2 \varphi(x,z,t)}{\partial z^2} - \frac{1}{c^2} \frac{\partial^2 \varphi(x,z,t)}{\partial t^2} =+k_c^2 \varphi(x,z,t)  , \label{veloceTE2}
\end{equation}
and similarly for $\varphi'$. Both fluxes propagate according to the Klein-Gordon equation, with the same mass term  $k_c \, \hbar /c = m  $.
Inspecting Eq. (\ref{enintTE2}), we find out that the energy densities can be written in the form:
\begin{eqnarray}
H_d & = & \frac{1}{2}\, C_d^{-1} \,  \sigma_t^{ 2} + \frac{1}{2} \, L_d  \, \vec{j}_t^{\, 2} - \frac{1}{4} \, L_d  \, j_{t\, x}^{\, 2}, \\
H_d' & = & \frac{1}{2}\, C_d\,\!\!'^{-1} \,  \sigma_l^{ 2} + \frac{1}{2} \, L_d\,\!\!'  \, \vec{j}_l^{\, 2} - \frac{1}{4} \, L_d\,\!\!'  \, j_{l\,y}^{\, 2} ,
\end{eqnarray}
or equivalently with bottom and right contributions. These are the usual charge/current energy densities, with an extra term arising from the symmetric splitting of the volumic integral over the $B_z^2$ field component. Note the difference with Eq. (\ref{HdTE}). \\

This section enabled us to derive the properties of the light field from the generalized fluxes.
Their propagation equation depends on the type of the wave, and leads to different interpretations for the nature of the confined photons: this is summarized in Tab. \ref{tab_bis}.
After performing the integrals in the expressions of energy $H$ and momentum $\vec{P}$ provided here, we will quantize the field quadratures $X, Y$ in Section \ref{quantize}; energy addenda and photon mass will also be discussed. For now, the next Section will link the generalized 
fluxes $\varphi, \varphi'$ with the electromagnetic potentials.

\begin{table}[h!]
\center
\caption{Propagation equation for $\varphi$ depending on the wave type.  }
\begin{tabular}{@{}lll}
\br
Propagation equation & Wave type & Photon property \\
\mr
$\frac{\partial^2 \varphi(x,z,t)}{\partial z^2} - \frac{1}{c^2}  \frac{\partial^2 \varphi(x,z,t)}{\partial t^2} =0$ & TEM wave & no mass, no addendum \\
$\frac{\partial^2 \varphi(x,z,t)}{\partial z^2} - \frac{1}{c^2} \left(\frac{\beta}{k}\right)^{\!\!2} \frac{\partial^2 \varphi(x,z,t)}{\partial t^2} =0$ & TM wave & energy addendum \\
$\frac{\partial^2 \varphi(x,z,t)}{\partial z^2} - \frac{1}{c^2} \frac{\partial^2 \varphi(x,z,t)}{\partial t^2} =+k_c^2 \varphi(x,z,t) $ & TE wave & photon mass \\
\br
\label{tab_bis}
\end{tabular}
\end{table}
\normalsize

\section{Electromagnetic gauge}

An important feature of Maxwell's equations is that the physical fields $\vec{E},\vec{B}$ can be derived from two other {\it potential fields} \cite{cohen}:
\begin{eqnarray}
\vec{E}(\vec{r},t) & = & -\frac{\partial \vec{A}(\vec{r},t)}{\partial t} - \vec{\mbox{grad}} \,V(\vec{r},t), \\
\vec{B}(\vec{r},t) & = & + \vec{\mbox{rot}} \,\vec{A}(\vec{r},t), 
\end{eqnarray}
with $V(\vec{r},t)$ the scalar potential and $\vec{A}(\vec{r},t)$ the vector potential.
It turns out from these expressions that the electromagnetic field is unchanged if one preforms the replacement:
\begin{eqnarray}
\vec{A}(\vec{r},t) & \rightarrow & \vec{A}(\vec{r},t) + \vec{\mbox{grad}} \, \Pi(\vec{r},t), \\
V(\vec{r},t) & \rightarrow & V(\vec{r},t) -\frac{\partial \, \Pi(\vec{r},t)}{\partial t}, 
\end{eqnarray}
where $\Pi(\vec{r},t)$ is an arbitrary function called {\it gauge}. This feature is what is named {\it gauge invariance}, a concept that appears also in other contexts \cite{rovelli}. As discussed in this Reference, modern physics considers that this is a fundamental property of Nature; we discuss it below in our electromagnetic context. \\

Actually, not any function $\Pi$ is relevant to a given problem. Obviously for us, we should consider only gauges that involve the degree of freedom $X,Y$ which characterizes the propagating field. As such, we should have:
\begin{equation}
\Pi(\vec{r},t) = p(x,y) f(z,t)+ \tilde{p}(x,y) \tilde{f}(z,t). \label{gauge}
\end{equation}
Any other function is essentially a {\it trivial invariance}, with no physical importance.
Besides, the gauge should share the symmetries of the field it applies to; this is already why in Eq. (\ref{gauge}), the quality of propagating solution led us to factorize the transverse $x,y$ dependence from the $z,t$ one.  
An important consequence of this argument is that the gauge should fullfill Lorentz invariance, that is it must comply with special relativity as the $\vec{E},\vec{B}$ fields do.
We therefore impose:
\begin{equation}
\mbox{div} \vec{A}(\vec{r},t) + \frac{1}{c^2} \frac{\partial V(\vec{r},t) }{\partial t} =0, \label{lorenz}
\end{equation}
which is known as the {\it Lorenz gauge condition} \cite{cohen}. Injecting Eq. (\ref{gauge}) into Eq. (\ref{lorenz}) leads to:
\begin{eqnarray}
\frac{\partial^2 p(x,y)}{\partial x^2} +  \frac{\partial^2 p(x,y)}{\partial y^2} + \left(k^2 - \beta^2 \right) p(x,y) &=&0, \label{pi} \\
\frac{\partial^2 \tilde{p}(x,y)}{\partial x^2} +  \frac{\partial^2 \tilde{p}(x,y)}{\partial y^2} + \left(k^2 - \beta^2 \right) \tilde{p}(x,y) &=&0  . \label{pi2}
\end{eqnarray}
The condition can effectively be split in two (equivalent) equations for $p$ and $\tilde{p}$ independently, since the functions $f,\tilde{f}$ are themselves orthogonal. We shall solve them for each specific case, taking into account the spatial transverse symmetries. \\

\subsection{Parallel plates}

For TEM waves, $ k = |\beta|$ and the functions are independent of $x$. Solving Eqs. (\ref{pi},\ref{pi2}) gives straightforwardly:
\begin{equation}
p(x,y) = a_\pi \, y + b_\pi , 
\end{equation}
with $a_\pi, b_\pi$ two unknowns, and an expression alike with tilded quantities.   
The formalism of the previous Section that involves $\varphi$ matches what is found in the literature, as it should \cite{ClerckDevoret,valraff}. It is transposed into voltage differences $\Delta V$ and currents $I$ in order to introduce the language of conventional electronics; this makes the link with the telegrapher's equation for signal transport, and circuit theory for lumped elements. These quantities can then be quantized, following an adequate procedure, in order to describe quantum circuits \cite{devoret97}. A key relationship introduced in this Reference is:
\begin{equation}
\frac{\partial \varphi}{\partial t} = \Delta V ,\label{deltaV}
\end{equation}
with $\Delta V = V(x,d/2,z,t)-V(x,-d/2,z,t)$ for a parallel plate guide, the {\it voltage difference} between top and bottom electrodes (the $x$ variable being irrelevant here). 
This formula deserves to be commented on: {\it it is not gauge invariant}. 
It actually relies on the fact that there must be a specific gauge which makes Eq. (\ref{deltaV}) meaningful. Precisely, we will demonstrate below that this assumption also applies to the other types of traveling waves (with some adaptation). The claim is essentially that, starting from the initial gauge invariance, we must fix (partially or totally) the gauge to obtain a result of the type of Eq. (\ref{deltaV}). This is precisely what is named {\it gauge fixing} in Ref. \cite{rovelli}. If part of the gauge is still unknown, this is its true {\it gauge invariance} (as opposed to the trivial one). Note that a direct consequence of the Lorenz invariance is that:
\begin{equation}
\frac{\partial \varphi}{\partial z} = -\Delta A ,\label{deltaA}
\end{equation}
having defined $\Delta A = A_z(x,d/2,z,t)-A_z(x,-d/2,z,t)$.

The TEM wave results are summarized in Tab. \ref{tab_9}. 
The components of $\vec{A}, V$ are listed, with the gauge coefficients $a_\pi, b_\pi, \tilde{a}_\pi, \tilde{b}_\pi$. 
The gauge fixing has been chosen such that fixed coefficients are set to zero; the others remain unknown. We recognize the usual {\it longitudinal gauge}. The same structure in the formula typesetting will be used for the other configurations, for reasons of readability. \\

\begin{table}[h!]
\center
\caption{Gauge description for TEM waves. }
\begin{tabular}{@{}lll}
\br
Function & & Expression \\
\mr
$A_x (x,y,z,t) $ & $=$ & $0$ \\
$A_y (x,y,z,t) $ & $=$ & $0$ \\
$A_z (x,y,z,t) $ & $=$ & $ +\phi_m \, \beta\, \left( \, \frac{y}{d} \, \right) \, f(z,t) $ \\
$V (x,y,z,t) $   & $=$ & $ +\phi_m \, \omega\, \left( \,\frac{y}{d}\, \right) \, f(z,t)$ \\
\mr
Gauge fixing & & $a_\pi =0$, $\tilde{a}_\pi =0$ \\
Gauge invariance  &  & $b_\pi$, $\tilde{b}_\pi $ \\
\br
\label{tab_9}
\end{tabular}
\end{table}
\normalsize

Let us now apply the same ideas to TM$_n$ waves. Here $k^2 - \beta^2 =k_c^2$, and the gauge must still be independent of $x$. This leads to:
\begin{equation}
p(x,y) = a_\pi \, \cos[k_c(y+d/2)] + b_\pi \, \sin[k_c(y+d/2)] , 
\end{equation}
and similarly for tilded parameters. Eqs. (\ref{deltaV}) and (\ref{deltaA}) can be generalized if we define:
\begin{eqnarray}
\Delta V & = & V(x,d/2,z,t)+(-1)^n \, V(x,-d/2,z,t) , \\
\Delta A & = & A_z(x,d/2,z,t)+(-1)^n \, A_z(x,-d/2,z,t) .
\end{eqnarray}
For profiles which have {\it symmetric} transverse components in the $\vec{y}$ direction (i.e. for $n$ even), we must {\it add up} the potentials present on the electrodes. For {\it anti-symmetric} modes ($n$ odd), we take the difference.   
The results are summarized in Tab. \ref{tab_10}. \\

\begin{table}[h!]
\center
\caption{Gauge description for TM$_n$ waves.  $n  \in \N^*$. }
\begin{tabular}{@{}lll}
\br
Function & & Expression \\
\mr
$A_x (x,y,z,t) $ & $=$ & $0$ \\
$A_y (x,y,z,t) $ & $=$ & $+ \phi_m \, (-1)^n \, \beta \, \left( \, \frac{1}{d\, \beta} \cos[k_c(y+d/2)] + \frac{k_c}{2\beta} \sin [k_c(y+d/2)] \, \right) \tilde{f}(z,t)  $ \\
$A_z (x,y,z,t) $ & $=$ & $+ \phi_m \, (-1)^n \,\beta \, \left( \, \frac{1}{2} \cos[k_c(y+d/2)] + \frac{2 k_c^2 +\beta^2}{d\, k_c \,\beta^2} \sin [k_c(y+d/2)] \, \right)  f (z,t)$ \\
$V (x,y,z,t) $   & $=$ & $+ \phi_m \, (-1)^n \,\omega\, \left( \, \frac{1}{2} \cos[k_c(y+d/2)] + \frac{1}{d\, k_c } \sin [k_c(y+d/2)] \, \right)  f (z,t)$ \\
\mr
Gauge fixing & & $a_\pi =0$, $\tilde{a}_\pi =0$ \\
Gauge invariance  &  & $b_\pi$, $\tilde{b}_\pi $ \\
\br
\label{tab_10}
\end{tabular}
\end{table}
\normalsize

The TE$_n$ traveling solutions are treated the same way, but this time we must use the lateral (virtual) electrodes. The gauge functions read the same as for TM$_n$ waves:
\begin{equation}
p(x,y) = a_\pi \, \cos[k_c(y+d/2)] + b_\pi \, \sin[k_c(y+d/2)] , \label{gaugepTE}
\end{equation}
and similarly for tilded quantities. Again we have no $x$ dependence, but we keep the variable for clarity. The gauge relations will be written here:
\begin{eqnarray}
\frac{\partial \varphi'}{\partial t} & = & \Delta V' , \\ 
\frac{\partial \varphi'}{\partial z} & = & - \Delta A' , \\
\Delta V' & = & V(w/2,y,z,t)+ V(-w/2,y,z,t) , \\
\Delta A' & = & A_z(w/2,y,z,t)+ A_z(-w/2,y,z,t) ,
\end{eqnarray}
the prime reminding that we are dealing with left/right boundaries. In the above, we must add the potentials present on each electrode for {\it any} $n$;
 the symmetry of the electromagnetic field profile in the $\vec{x}$ direction does not depend on the index $n$ (it is actually always symmetric, since independent of $x$).
The obtained formulas are given 
in Tab. \ref{tab_11}. Note that for this type of traveling wave, the gauge is completely fixed.

\begin{table}[h!]
\center
\caption{Gauge description for TE$_n$ waves. $n  \in \N^*$. }
\begin{tabular}{@{}lll}
\br
Function & & Expression \\
\mr
$A_x (x,y,z,t) $ & $=$ & $+\phi_m' \, \beta \, \left( \,\frac{1}{w\, \beta} \sin [k_c(y+d/2)] \,\right)  \tilde{f}(z,t) $ \\
$A_y (x,y,z,t) $ & $=$ & $-\phi_m' \, \beta \,\left( \,\frac{k_c}{2\beta} \cos [k_c(y+d/2)] \,\right)  \tilde{f}(z,t)$ \\
$A_z (x,y,z,t) $ & $=$ & $+\phi_m' \, \beta \, \left( \,\frac{1}{2} \sin [k_c(y+d/2)] \,\right)   f(z,t)$ \\
$V (x,y,z,t) $   & $=$ & $+\phi_m' \, \omega \, \left( \,\frac{1}{2} \sin [k_c(y+d/2)] \,\right)   f (z,t)$ \\
\mr
Gauge fixing & & $a_\pi =0$, $b_\pi=0$, $\tilde{a}_\pi =0$, $\tilde{b}_\pi =0$ \\
Gauge invariance  &  &  None \\
\br
\label{tab_11}
\end{tabular}
\end{table}
\normalsize
\vspace*{3mm} 

\subsection{Rectangular guide}

\begin{table}[t!]
\center
\caption{Gauge description for TM$_{n,m}$ waves. $n,m  \in \N^*$. }
\hspace*{-10mm}
\begin{tabular}{@{}lll}
\br
Function & & Expression \\
\mr
$A_x (x,y,z,t) $ & $=$ & $+(-1)^n\, \phi_m \, \beta \left( \, -\frac{k_{cx}}{2\beta }  \cos[k_{cx}(x+w/2)] \cos[k_{cy}(y+d/2)]\, \right) \tilde{f}(z,t) $ \\
&     & $\!\!\!\!\!\!\!\!\!\!\!\!\!\!\!\!\!\!\!\!\!\!\!\!\!\!\!\!\!\!\!\!\!\!\!\!\!\!\!\!\!\!\!\!\!\!\!\!\!\!\!\!\!\!\!\!\!\! +(-1)^m\, \phi_m' \, \beta \left( \,\frac{4}{w\,\beta}  \cos[k_{cx}(x+w/2)]\sin[k_{cy}(y+d/2)]+\frac{k_{cx}}{2\beta }  \sin[k_{cx}(x+w/2)]\sin[k_{cy}(y+d/2)] \, \right) \tilde{f}(z,t) $ \\
$A_y (x,y,z,t) $ & $=$ & $+(-1)^n\, \phi_m \, \beta \left( \, +\frac{k_{cy}}{2\beta }  \sin[k_{cx}(x+w/2)] \sin[k_{cy}(y+d/2)]\, \right) \tilde{f}(z,t)$ \\
&    & $\!\!\!\!\!\!\!\!\!\!\!\!\!\!\!\!\!\!\!\!\!\!\!\!\!\!\!\!\!\!\!\!\!\!\!\!\!\!\!\!\!\!\!\!\!\!\!\!\!\!\!\!\!\!\!\!\!\! +(-1)^m\, \phi_m' \, \beta \left( \,\frac{4 k_{cy}}{w\,k_{cx} \, \beta}  \sin[k_{cx}(x+w/2)]\cos[k_{cy}(y+d/2)]-\frac{k_{cy} }{2\beta }  \cos[k_{cx}(x+w/2)]\cos[k_{cy}(y+d/2)] \, \right) \tilde{f}(z,t) $ \\
$A_z (x,y,z,t) $ & $=$ & $+(-1)^n\, \phi_m \, \beta \left( \, \frac{1}{2}  \sin[k_{cx}(x+w/2)] \cos[k_{cy}(y+d/2)]\, \right) f(z,t) $ \\
&     & $\!\!\!\!\!\!\!\!\!\!\!\!\!\!\!\!\!\!\!\!\!\!\!\!\!\!\!\!\!\!\!\!\!\!\!\!\!\!\!\!\!\!\!\!\!\!\!\!\!\!\!\!\!\!\!\!\!\! +(-1)^m\, \phi_m' \, \beta \left( \,\frac{1}{2}  \cos[k_{cx}(x+w/2)]\sin[k_{cy}(y+d/2)]+\frac{2 k_c^2-2 \beta^2}{w\, k_{cx}\,\beta^2}  \sin[k_{cx}(x+w/2)]\sin[k_{cy}(y+d/2)] \, \right) f(z,t) $ \\
$V(x,y,z,t)$       & = & $+(-1)^n\, \phi_m \, \omega \left( \, \frac{1}{2}  \sin[k_{cx}(x+w/2)] \cos[k_{cy}(y+d/2)]\, \right) f(z,t) $ \\
                 &     & $\!\!\!\!\!\!\!\!\!\!\!\!\!\!\!\!\!\!\!\!\!\!\!\!\!\!\!\!\!\!\!\!\!\!\!\!\!\!\!\!\!\!\!\!\!\!\!\!\!\!\!\!\!\!\!\!\!\! +(-1)^m\, \phi_m' \, \omega \left( \,\frac{1}{2} \cos[k_{cx}(x+w/2)]\sin[k_{cy}(y+d/2)] - \frac{2}{w \, k_{cx}} \sin[k_{cx}(x+w/2)]\sin[k_{cy}(y+d/2)]  \,\right) f(z,t)$ \\
\mr
Gauge fixing & & $a_\pi =0$,  $\tilde{a}_\pi =0$, $c_\pi =0$,  $\tilde{c}_\pi =0$, $d_\pi =0$,  $\tilde{d}_\pi =0$, \\
Gauge invariance  &  &  $b_\pi$, $\tilde{b}_\pi$\\
\br
\label{tab_12}
\end{tabular}
\end{table}
\normalsize

The symmetries of TM$_{n,m}$ traveling modes lead to the gauge functions:
\begin{eqnarray}
 \!\!\!\!\!\!\!\!\!\!\!\!\!\!\!\!\!\!\!\!\!\!\!\!\!\!\!\!\!\!\!\!\!\!\!\!\!\! p(x,y) &\!\!\!\!\!\!\!\!\!\!\!\!\!\!\!\!\!\!\!\! = &\!\!\!\!\!\!\!\!\!\!\! a_\pi \sin[k_{cx}(x+w/2)] \cos[k_{cy}(y+d/2)] + b_\pi \sin[k_{cx}(x+w/2)] \sin[k_{cy}(y+d/2)] \nonumber \\
&&\!\!\!\!\!\!\!\!\!\!\!\!\!\!\!\!\!\!\!\!\!\!\!\!\!\!\!\!\!\!\!\! + c_\pi \cos[k_{cx}(x+w/2)] \sin[k_{cy}(y+d/2)] + d_\pi \cos[k_{cx}(x+w/2)] \cos[k_{cy}(y+d/2)] , \label{gaugepTM}
\end{eqnarray}
and the equivalent expression for $\tilde{p}(x,y)$. The aim is now to find a gauge that fulfills {\it at the same time}:
\begin{eqnarray}
\frac{\partial \varphi}{\partial t} & = & \Delta V , \label{finalphi1} \\ 
\frac{\partial \varphi}{\partial z} & = & - \Delta A , \\
\frac{\partial \varphi'}{\partial t} & = & \Delta V' , \\ 
\frac{\partial \varphi'}{\partial z} & = & - \Delta A' , \label{finalphi4}
\end{eqnarray}
on the top/bottom and left/right electrodes, having defined: 
\begin{eqnarray}
\Delta V & = & V(x,d/2,z,t)+(-1)^n \, V(x,-d/2,z,t) , \\
\Delta A & = & A_z(x,d/2,z,t)+(-1)^n \, A_z(x,-d/2,z,t) , \\
\Delta V' & = & V(w/2,y,z,t)+(-1)^m \,  V(-w/2,y,z,t) , \\
\Delta A' & = & A_z(w/2,y,z,t)+(-1)^m\,  A_z(-w/2,y,z,t) . \label{Aprimeval}
\end{eqnarray}
The symmetry of the transverse electromagnetic field is now in-built in these expressions for the two directions of space $\vec{x}, \vec{y}$ with $n,m \in \N^*$.
The components of $\vec{A}, V$ fulfilling these conditions are given in Tab. \ref{tab_12}.
We remind that the chosen writing, which introduces both $\phi_m$ and $\phi_m'$, is a convenience which highlights the presence of the two sets of boundaries: there is physically only one degree of freedom in the problem (the $X,Y$ quadratures hidden in $f, \tilde{f}$). 

The TE$_{n,m=0}$ modes being strictly equivalent to the TE$_n$ ones, the corresponding potentials have been already given in Tab. \ref{tab_11}. The $90^\circ$-rotated solution TE$_{n=0,m}$ is straightforwardly obtained; the result is summarized in Tab. \ref{tab_13}. Note that in the gauge functions $p, \tilde{p}$ of Eq. (\ref{gaugepTE}) the variable dependence is modified by $y \rightarrow x, d \rightarrow w$ (and $k_c = k_{cx}$). \\

\begin{table}[h!]
\center
\caption{Gauge description for TE$_{n=0,m}$ waves. $m  \in \N^*$. }
\begin{tabular}{@{}lll}
\br
Function & & Expression \\
\mr
$A_x (x,y,z,t) $ & $=$ & $-\phi_m  \, \beta \, \left( \,\frac{k_{cx}}{2 \beta} \cos [k_{cx}(x+w/2)] \,\right)  \tilde{f}(z,t) $ \\
$A_y (x,y,z,t) $ & $=$ & $+\phi_m \, \beta \,\left( \,\frac{1}{d \, \beta} \sin [k_{cx}(x+w/2)] \,\right)  \tilde{f}(z,t)$ \\
$A_z (x,y,z,t) $ & $=$ & $+\phi_m \, \beta \, \left( \,\frac{1}{2} \sin [k_{cx}(x+w/2)] \,\right)   f(z,t)$ \\
$V (x,y,z,t) $   & $=$ & $+\phi_m \, \omega \, \left( \,\frac{1}{2} \sin [k_{cx}(x+w/2)] \,\right)   f (z,t)$ \\
\mr
Gauge fixing & & $a_\pi =0$, $b_\pi=0$, $\tilde{a}_\pi =0$, $\tilde{b}_\pi =0$ \\
Gauge invariance  &  &  None \\
\br
\label{tab_13}
\end{tabular}
\end{table}
\normalsize

Finally, the gauge functions $p(x,y)$ and $\tilde{p}(x,y)$ of the TE$_{n,m}$ waves with $n,m \in \N^*$ read the same as for the TM$_{n,m}$ modes, Eq. (\ref{gaugepTM}). Applying exactly the same equations linking $\varphi, \varphi'$ to the potentials $\vec{A}, V$, Eqs. (\ref{finalphi1}-\ref{Aprimeval}), we obtain the gauge expressions presented in 
Tab. \ref{tab_14}. \\

To conclude the Section, it appears that we can use 
Eqs. (\ref{finalphi1}-\ref{finalphi4}) to extend the definition of generalized 
flux, introduced originally in Ref. \cite{devoret97}, to {\it all} traveling waves treated here. We must replace the usual voltage difference by the quantities:
\begin{eqnarray}
\Delta V & = & V(x,d/2,z,t)+\sigma \, V(x,-d/2,z,t) , \\
\Delta A & = & A_z(x,d/2,z,t)+\sigma \, A_z(x,-d/2,z,t) , \\
\Delta V' & = & V(w/2,y,z,t)+\sigma' \,  V(-w/2,y,z,t) , \\
\Delta A' & = & A_z(w/2,y,z,t)+\sigma' \,  A_z(-w/2,y,z,t) ,
\end{eqnarray}
for top/bottom and left/right (real or virtual) electrodes (for the latter case, only if they are meaningful, see Section \ref{Max}).
The "parity" coefficients $\sigma,\sigma' = \pm 1$ depend on the symmetry of the transverse field, and the {\it nature} of the wave: 
for TEM (which is symmetric in the transverse plane) we have the usual $\sigma=-1$, while for all others a symmetric transverse profile corresponds to a  $\sigma,\sigma' = +1$ (and for anti-symmetric to a $-1$).
 Note that the non-trivial gauge invariance is actually quite limited, with only a couple of coefficients undefined at most.

\begin{table}[h!]
\center
\caption{Gauge description for TE$_{n,m}$ waves. $n,m  \in \N^*$. }
\hspace*{-10mm}
\begin{tabular}{@{}lll}
\br
Function & & Expression \\
\mr
$A_x (x,y,z,t) $ & $=$ & $+(-1)^n\, \phi_m \, \beta \left( \, -\frac{k_{cx}}{2\beta}  \cos[k_{cx}(x+w/2)] \cos[k_{cy}(y+d/2)]\, \right) \tilde{f}(z,t) $ \\
&     & $\!\!\!\!\!\!\!\!\!\!\!\!\!\!\!\!\!\!\!\!\!\!\!\!\!\!\!\!\!\!\!\!\!\!\!\!\!\!\!\!\!\!\!\!\!\!\!\!\!\!\!\!\!\!\!\!\!\!\!\!\!\!\!\!\!\! +(-1)^m\, \phi_m' \, \beta \left( \,\frac{2k_{cx}^2 + 2 k_{cy} \,\beta}{w\,k_{cy}\,\beta^2}  \cos[k_{cx}(x+w/2)]\sin[k_{cy}(y+d/2)]+\! \frac{k_{cx}}{2\beta }  \sin[k_{cx}(x+w/2)]\sin[k_{cy}(y+d/2)] \, \right) \tilde{f}(z,t) $ \\
$A_y (x,y,z,t) $ & $=$ & $+(-1)^n\, \phi_m \, \beta \left( \, +\frac{k_{cy}}{2\beta}  \sin[k_{cx}(x+w/2)] \sin[k_{cy}(y+d/2)]\, \right) \tilde{f}(z,t)$ \\
&  & $\!\!\!\!\!\!\!\!\!\!\!\!\!\!\!\!\!\!\!\!\!\!\!\!\!\!\!\!\!\!\!\!\!\!\!\!\!\!\!\!\!\!\!\!\!\!\!\!\!\!\!\!\!\!\!\!\!\!\!\!\!\!\!\!\!\! +(-1)^m\, \phi_m' \, \beta \left( \, \frac{2k_{cx}(k_{cy}- \beta)}{w \,k_{cy}\, \beta^2 }  \sin[k_{cx}(x+w/2)]\cos[k_{cy}(y+d/2)] -\! \frac{k_{cy}}{2\beta }  \cos[k_{cx}(x+w/2)]\cos[k_{cy}(y+d/2)] \, \right) \! \tilde{f}(z,t) $ \\
$A_z (x,y,z,t) $ & $=$ & $+(-1)^n\, \phi_m \, \beta \left( \, \frac{1}{2}  \sin[k_{cx}(x+w/2)] \cos[k_{cy}(y+d/2)]\, \right) f(z,t) $ \\
&     & $\!\!\!\!\!\!\!\!\!\!\!\!\!\!\!\!\!\!\!\!\!\!\!\!\!\!\!\!\!\!\!\!\!\!\!\!\!\!\!\!\!\!\!\!\!\!\!\!\!\!\!\!\!\!\!\!\!\!\!\!\!\!\!\! +(-1)^m\, \phi_m' \, \beta \left( \,\frac{1}{2}  \cos[k_{cx}(x+w/2)]\sin[k_{cy}(y+d/2)]-\frac{2 k_{cx}}{w\, k_{cy}\,\beta }  \sin[k_{cx}(x+w/2)]\sin[k_{cy}(y+d/2)] \, \right) f(z,t) $ \\
$V (x,y,z,t) $   & $=$ & $+(-1)^n\, \phi_m \, \omega \left( \, \frac{1}{2}  \sin[k_{cx}(x+w/2)] \cos[k_{cy}(y+d/2)]\, \right) f(z,t) $ \\
                 &     & $\!\!\!\!\!\!\!\!\!\!\!\!\!\!\!\!\!\!\!\!\!\!\!\!\!\!\!\!\!\!\!\!\!\!\!\!\!\!\!\!\!\!\!\!\!\!\!\!\!\!\!\!\!\!\!\!\!\!\!\!\!\!\!\! +(-1)^m\, \phi_m' \, \omega \left( \,\frac{1}{2} \cos[k_{cx}(x+w/2)]\sin[k_{cy}(y+d/2)] - \frac{2 k_{cx}}{w \, k_{cy}\, \beta} \sin[k_{cx}(x+w/2)]\sin[k_{cy}(y+d/2)]  \,\right) f(z,t)$ \\
\mr
Gauge fixing & & $a_\pi =0$,  $\tilde{a}_\pi =0$, $c_\pi =0$,  $\tilde{c}_\pi =0$, $d_\pi =0$,  $\tilde{d}_\pi =0$, \\
Gauge invariance  &  &  $b_\pi$, $\tilde{b}_\pi$\\
\br
\label{tab_14}
\end{tabular}
\end{table}
\normalsize

\section{Field quantization}
\label{quantize}

Before properly quantizing $X$ and $Y$ (our degree of freedom), consider the flux amplitudes:
\begin{eqnarray}
\varphi_{max}(z,t) &= & \phi_m \, \tilde{f}(z,t) , \\
\varphi_{max}'(z,t) &= & \phi_m' \, \tilde{f}(z,t) ,
\end{eqnarray}
which correspond to Eqs. (\ref{varphig1},\ref{varphig2}) without the $g_\phi(x), g_{\phi'}(y)$ position dependencies. 
The integration of the $H$ and $\vec{P}$ expressions over the transverse variables $x$ and $y$ brings:
\begin{eqnarray}
H & = &\!\!\! \int_{z=0}^L  \left[ \frac{1}{2}\, C_H^{-1} \left( C_H \frac{\partial \varphi_{max}(z,t)}{\partial t}\right)^{\!\!2} + \frac{1}{2} \, L_H^{-1} \left(  \frac{\partial \varphi_{max}(z,t)}{\partial z}\right)^{\!\!2} \right. \nonumber \\
&& \left. + \frac{1}{2}\, C_P \, (c\, k_c)^2 \,  \varphi_{max}(z,t)^{2}  \right]   dz/L , \label{finalHz} \\
\vec{P} & = &\!\!\!  \int_{z=0}^L     \left[\left( C_P \frac{\partial \varphi_{max}(z,t)}{\partial t} \right)\, \left( -  \frac{\partial \varphi_{max}(z,t)}{\partial z} \right) \right]  dz/L \, \vec{z} , \label{finalPz}
\end{eqnarray}
and the equivalent primed equations (for real or virtually defined lateral electrodes), for {\it all configurations} addressed here. The introduced coefficients, total capacitances $C_H, C_P$, and total inductance $L_H$ are listed in Tab. \ref{tab_15}. 
$C_H$ is the capacitance that appears in the "kinetic" term of the energy, while $C_P$ defines the momentum $\vec{P}$; it also shows up in the energy addendum (which Section \ref{charges} identified as being of kinetic origin for TE waves, and potential for TM; obviously for TEM modes $k_c=0$).  
For TEM, TM$_n$, TE$_n$, and TE$_{n=0,m}$ traveling waves,
{\it only one} pair of electrodes is meaningful and required for the description of the electromagnetic field. 
With TM$_{n,m}$ and TE$_{n,m}$ solutions having $n,m \in \N^*$ however, top/bottom and left/right electrodes play an equivalent role:
both $H$ and $\vec{P}$ can be expressed similarly with either non-primed or primed quantities.
Tab. \ref{tab_15} also reminds the parameters $h_{e\!f\!f},h_{e\!f\!f}'$ and $\sigma,\sigma'$ for completeness, for all possible cases.  Eqs. (\ref{finalHz},\ref{finalPz}) lead us to define a new important quantity:
\begin{equation}
Q_{max}(z,t) = +C_p  \, \frac{\partial \varphi_{max}(z,t)}{\partial t} ,
\end{equation}
for top/bottom electrodes, and similarly for left/right with primes. This corresponds to a {\it charge amplitude} (in Coulomb), displaying a propagative $z,t$ dependence through $\propto f(z,t)$. \\
 
\begin{table}[h!]
\center
\caption{Modal coefficients $C_H, C_P$ and $L_H$ (together with $h_{e\!f\!f}$, $\sigma$, $k_c$ and phase velocity $v_\phi$). We remind that $C_d=\epsilon_0/h_{e\!f\!f}$ and $L_d^{-1}=1/(\mu_0\,h_{e\!f\!f})$, $k_{cx}= m \pi/w$ and $k_{cy}= n \pi/d$ with $n,m \in \N^*$. And similarly for primed quantitites.  }
\begin{tabular}{@{}lll}
\br
Wave type & Modal coefficients &  Potential related parameters \\
\mr
TEM & $C_H = C_d \, w \, L$ &  $k_c=0$; velocity $c$ \\
    & $C_P = C_H $ &   $h_{e\!f\!f}=d$        \\
    & $L_H^{-1} =L_d^{-1} \, w \, L $ & $\sigma=-1$     \\
\mr
TM$_n$ & $C_H= C_d \, w \, L$ &   $k_c=k_{cy}$; velocity $v_\phi=c \, k/|\beta|$   \\
    & $C_P= C_H \left( k/\beta \right)^2 $ &  $h_{e\!f\!f}=d/2$       \\
    & $L_H^{-1}=L_d^{-1} \, w \, L \left( k/\beta \right)^4 $ & $\sigma=(-1)^n$     \\
\mr
TE$_n$ & $C_H'= C_d' \, (d/2) \, L $ & $k_c=k_{cy}$; velocity $c$  \\
(or TE$_{n,m=0}$)    & $C_P'=C_H'$ &    $h_{e\!f\!f}'=w$      \\
    & $L_H'^{-1} =L_d'^{-1} \, (d/2) \, L$ & $\sigma'=+1$     \\
\mr
TE$_{n=0,m}$ & $C_H=C_d \, (w/2) \, L $ & $k_c=k_{cx}$; velocity $c$ \\
    & $C_P=C_H$ &  $h_{e\!f\!f}=d$         \\
    & $L_H^{-1}=L_d^{-1} \, (w/2) \, L$ & $\sigma=+1$     \\
\mr
\mr
TM$_{n,m}$ & $C_H= C_d \, (w/2)\,\left(1+k_{cx}^2/k_{cy}^2 \right) \, L$ &  $k_c=\sqrt{k_{cx}^2+k_{cy}^2}$; velocity $v_\phi=c \, k/|\beta|$    \\
top/bottom    & $C_P= C_H \left( k/\beta \right)^2$ & $h_{e\!f\!f}=d/2$        \\
    & $L_H^{-1} =L_d^{-1} \, (w/2)\,\left(1+ k_{cx}^2/k_{cy}^2\right) \, L \left( k/\beta \right)^4 $ & $\sigma=(-1)^n$     \\
\mr
TM$_{n,m}$ & $C_H'= C_d' \, (d/2)\,\left(1+k_{cy}^2/k_{cx}^2 \right) \, L$ &    $k_c=\sqrt{k_{cx}^2+k_{cy}^2}$; velocity $v_\phi=c \, k/|\beta|$   \\
left/right    & $C_P'= C_H' \left( k/\beta \right)^2$ &   $h_{e\!f\!f}'=w/2$       \\
    & $L_H'^{-1}=L_d'^{-1} \, (d/2)\,\left(1+ k_{cy}^2/k_{cx}^2\right) \, L \left( k/\beta \right)^4$ & $\sigma'=(-1)^m$     \\
\mr
TE$_{n,m}$ & $C_H= C_d \, (w/2)\,\left(1+ k_{cy}^2/k_{cx}^2\right) \, L$ &  $k_c=\sqrt{k_{cx}^2+k_{cy}^2}$; velocity $c$  \\
top/bottom     & $C_P=C_H$ &   $h_{e\!f\!f}=d/2$       \\
    & $L_H^{-1}=L_d^{-1} \, (w/2)\,\left(1+ k_{cy}^2/k_{cx}^2\right) \, L $ & $\sigma=(-1)^n$     \\
\mr
TE$_{n,m}$ & $C_H'= C_d' \, (d/2)\,\left(1+k_{cx}^2/k_{cy}^2\right) \, L$ &   $k_c=\sqrt{k_{cx}^2+k_{cy}^2}$; velocity $c$    \\
left/right    & $C_P'=C_H'$ &  $h_{e\!f\!f}'=w/2$     \\
    & $L_H'^{-1}=L_d'^{-1} \, (d/2)\,\left(1+ k_{cx}^2/k_{cy}^2 \right) \, L $ & $\sigma'=(-1)^m$     \\
\br
\label{tab_15}
\end{tabular}
\end{table} 
\normalsize 

We now promote the two quadratures to operators. As opposed to other approaches \cite{chineseStuff,roumains}, we do not aim at  linking this theoretical description to the conventional relativistic treatment of light (and especially to the {\it spin}), somehow justifying the quantumness of photons traveling in a waveguide. Neither do we want to explore speculative new quantum applications of traveling fields \cite{scalarphi,chiralW}. What we propose to do is much more practical, and aims at understanding {\it the consequences for the measurable fields} of the conventional condensed matter approach of quantum information transfer \cite{ClerckDevoret,valraff}, taking it as granted. We therefore pose:
\begin{eqnarray}
X & \longrightarrow & \hat{X} , \\
Y & \longrightarrow & \hat{Y} ,
\end{eqnarray}
the hat symbol denoting operators. Performing the last integral over $z$, we obtain:
\begin{eqnarray}
\hat{H} & = & ( 2 C_P \, \omega \, \phi_m^2 ) \,\omega \left( \frac{\hat{X}^2 + \hat{Y}^2}{4} \right) , \label{final1} \\
\hat{\vec{P}} & = & ( 2 C_P \, \omega \, \phi_m^2 ) \,\beta \left( \frac{\hat{X}^2 + \hat{Y}^2}{4} \right) \, \vec{z} ,
\end{eqnarray} 
while $\hat{Q}_{max}$ and $\hat{\varphi}_{max}$ verify the commutation rule:
\begin{equation}
\left[\hat{\varphi}_{max}(z,t),\hat{Q}_{max}(z,t) \right] = ( 2 C_P \, \omega \, \phi_m^2 )  \left[\hat{X},\hat{Y}\right]\!/\, 2  . \label{final3}
\end{equation}
Similar equations are obtained with $C_P'$ and $\phi_m'$, when these apply (lateral electrodes well defined). 
 Eqs. (\ref{final1}-\ref{final3}) are all position and time independent, and involve the same prefactor: $2 C_P \, \omega \, \phi_m^2 $ (or $2 C_P' \, \omega \, \phi_m'^2 $), in Joule.seconds. We finally introduce the {\it creation and annihilation operators} \cite{cohen,landau}:
\begin{eqnarray}
\hat{X} & = & +\left( \hat{b}^\dag + \hat{b} \right), \label{bbhat1} \\
\hat{Y} & = & +\mbox{i} \left( \hat{b}^\dag - \hat{b}\right) ,
\end{eqnarray}
which recast the $\hat{f}, \hat{\tilde{f}}$ functions into:
\begin{eqnarray}
\hat{f}(z,t)      & = & \left( \hat{b}^\dag e^{+\mbox{i} \, \theta_0 } \right) e^{+\mbox{i} \,(\omega t- \beta z) } + \left( \hat{b} \, e^{-\mbox{i} \, \theta_0 } \right) e^{-\mbox{i} \,(\omega t- \beta z) } , \label{bbhat3} \\
\hat{\tilde{f}}(z,t) & = &-\mbox{i} \left[  \left( \hat{b}^\dag e^{+\mbox{i} \, \theta_0 } \right) e^{+\mbox{i} \,(\omega t- \beta z) } - \left( \hat{b} \,  e^{-\mbox{i} \, \theta_0 } \right) e^{-\mbox{i} \,(\omega t- \beta z) } \right] ,  \label{bbhat4}
\end{eqnarray}
and bring the relationships:
\begin{eqnarray}
 \left[\hat{X},\hat{Y}\right]& =& +2 \mbox{i}  \left[\hat{b} ,\hat{b}^\dag\right] , \label{comXY} \\
 \hat{X}^2 + \hat{Y}^2 & = & +4 \left(\hat{b}^\dag \hat{b} + \frac{\left[\hat{b} ,\hat{b}^\dag\right]}{2} \right) . \label{X2Y2}
\end{eqnarray}
Eqs. (\ref{final1}-\ref{final3}) become our conventional textbook expressions if one imposes:
\begin{eqnarray}
\left[\hat{b} ,\hat{b}^\dag\right] & = & 1 , \label{bosons} \\
2 C_P \, \omega \, \phi_m^2 & = & \hbar , \label{hbarval}
\end{eqnarray}
or equivalently $2 C_P' \, \omega \, \phi_m'^2  =\hbar$ when dealing with lateral electrodes. The photon population operator is then defined as $\hat{n}=\hat{b}^\dag \hat{b} $, which can be replaced in Eq. (\ref{X2Y2}) \cite{cohen,landau}. The charge $\hat{Q}_{max}$ acquires now a very clear meaning: it is the {\it quantum conjugate} of the flux amplitude $\hat{\varphi}_{max}$ \cite{ClerckDevoret,valraff}, defined even for virtual electrode pairs.
Note that Eqs. (\ref{comXY},\ref{X2Y2}) read the same for time-shifted operators $\hat{b}_r = \hat{b} \, e^{-\mbox{i} \, \theta_0 }$ and $\hat{b}_r^\dag = \hat{b}^\dag e^{+\mbox{i} \, \theta_0 }$. These correspond to a {\it rotation} of the quadratures as $ \hat{X} = \cos(\theta_0) \, \hat{X}_r - \sin(\theta_0) \,  \hat{Y}_r$ and $ \hat{Y} = \sin(\theta_0) \, \hat{X}_r + \cos(\theta_0) \, \hat{Y}_r$. 
As well, changing the overall sign of {\it either} $\hat{X}$ {\it or} $\hat{Y}$  (i.e. mirror symmetry in the quadrature space) {\it changes the sign} of the commutator Eq. (\ref{comXY}), without altering the writing of $\hat{H}$ and $\hat{\vec{P}}$: it is equivalent to exchanging the roles of creation $\hat{b}^\dag$ and annihilation $\hat{b}$ operators in Eqs. (\ref{bbhat3},\ref{bbhat4}). 
Finally, the presented treatment is  invariant under an overall scaling by a real number $\alpha \neq 0$, such that $\phi_m \rightarrow \alpha \, \phi_m $, implying:
\begin{eqnarray}
\hat{\varphi}_{max} \rightarrow \alpha \, \hat{\varphi}_{max}, \\
\hat{Q}_{max} \rightarrow \hat{Q}_{max}/ \alpha, \\
C_{P,H} \rightarrow C_{P,H}/\alpha^2, \\
L_H \rightarrow L_H \, \alpha^2,
\end{eqnarray}
and equivalently with primed quantities. This is simply linked to our somewhat arbitrary choice of describing the collective mode by means of its maximal magnitude; a similar situation is encountered in mechanics, when defining an {\it effective mass} (role played here by a capacitance) assigned to a collective motion. It is our freedom when describing the problem as a whole, and has {\it nothing to do} with the elementary definitions of $C_d, L_d$ or $h_{ef\!f}$.
\\

Let us now comment Eq. (\ref{finalHz}). The energy integral can be split in two:
\begin{eqnarray}
\!\!\!\!\!\!\!\! \!\!\!\!\!\!\!\!\!\!\!\!\!\!  \hat{H} - \frac{\Delta \hat{H}}{2} &   = &   \!\!   \int_{z=0}^L  \left[ \frac{1}{2}\, C_H^{-1} \left( C_H \frac{\partial \hat{\varphi}_{max}(z,t)}{\partial t}\right)^{\!\!2} + \frac{1}{2} \, L_H^{-1} \left(  \frac{\partial \hat{\varphi}_{max}(z,t)}{\partial z}\right)^{\!\!2} \right]   dz/L  , \label{mainAd} \\
\!\!\!\!\!\!\!\! \!\!\!\!\!\!\!\! \!\!\!\!\!\! +\frac{\Delta \hat{H}}{2} &   = &   \!\!   \int_{z=0}^L  \left[ \frac{1}{2}\, C_P \, (c\, k_c)^2 \,  \hat{\varphi}_{max}(z,t)^{2}  \right]   dz/L  ,   \label{addH}
\end{eqnarray}
identifying the energy addendum (nonzero  for TE, TM modes $k_c \neq 0$):
\begin{equation}
\Delta \hat{H} =  \hbar\, \omega_c  \sqrt{\frac{k_c^2}{\beta^2+k_c^2} } \left( \hat{n} + \frac{1}{2}\right) ,
\end{equation}
with $\omega_c = c \, k_c$.
Adding a photon $\delta \! < \!\hat{n}\! >=+1$ in the waveguide thus costs an extra energy $\hbar\, \omega_c \sqrt{ k_c^2/(\beta^2+k_c^2)}$, which is split for half in the addendum term Eq. (\ref{addH}), and for half in the main term Eq. (\ref{mainAd}). 
As pointed out in Section \ref{charges}, for TM waves this corresponds to an extra potential energy, which we interpret as the cost of confining a photon in the traveling line. 
For TE waves, the origin is "kinetic", in the sense that the equations are of the Klein-Gordon type. The addendum can then be recast into a {\it photon mass} with:
\begin{equation}
 \hbar \, \omega_c = m\, c^2 ,
\end{equation} 
and corresponds to the {\it rest energy} of the massive photon. \\

\begin{table}[h!]
\center
\caption{Zero-point fluctuations $E_m$ (referenced to top/bottom electrodes) for all configurations. $n,m \in \N^*$. }
\begin{tabular}{@{}ll}
\br
Wave type & Zero-point fluctuation electric field  \\
\mr
TEM & $E_m = \sqrt{\frac{\hbar \, \omega/2 }{\epsilon_0 \, (d\, w\, L)}}= E_{zpf} $  \\
\mr
TM$_n$ & $E_m = \sqrt{2} \, E_{zpf} \sqrt{\frac{\beta^2}{\beta^2+k_c^2}}$   \\
\mr
TE$_{n=0,m}$ & $E_m =\sqrt{2} \, E_{zpf}  $ \\
\mr
\mr
TM$_{n,m}$ & $E_m = \sqrt{\frac{4}{1+k_{cx}^2/k_{cy}^2}} \, E_{zpf} \sqrt{\frac{\beta^2}{\beta^2+k_c^2}} $    \\
\mr
TE$_{n,m}$ & $E_m =\sqrt{\frac{4}{1+k_{cy}^2/k_{cx}^2}} \, E_{zpf} $  \\
\br
\label{tab_16}
\end{tabular}
\end{table}
\normalsize

\begin{figure}[t!]
	\vspace*{-1.1cm}
		\centering
	\includegraphics[width=28cm]{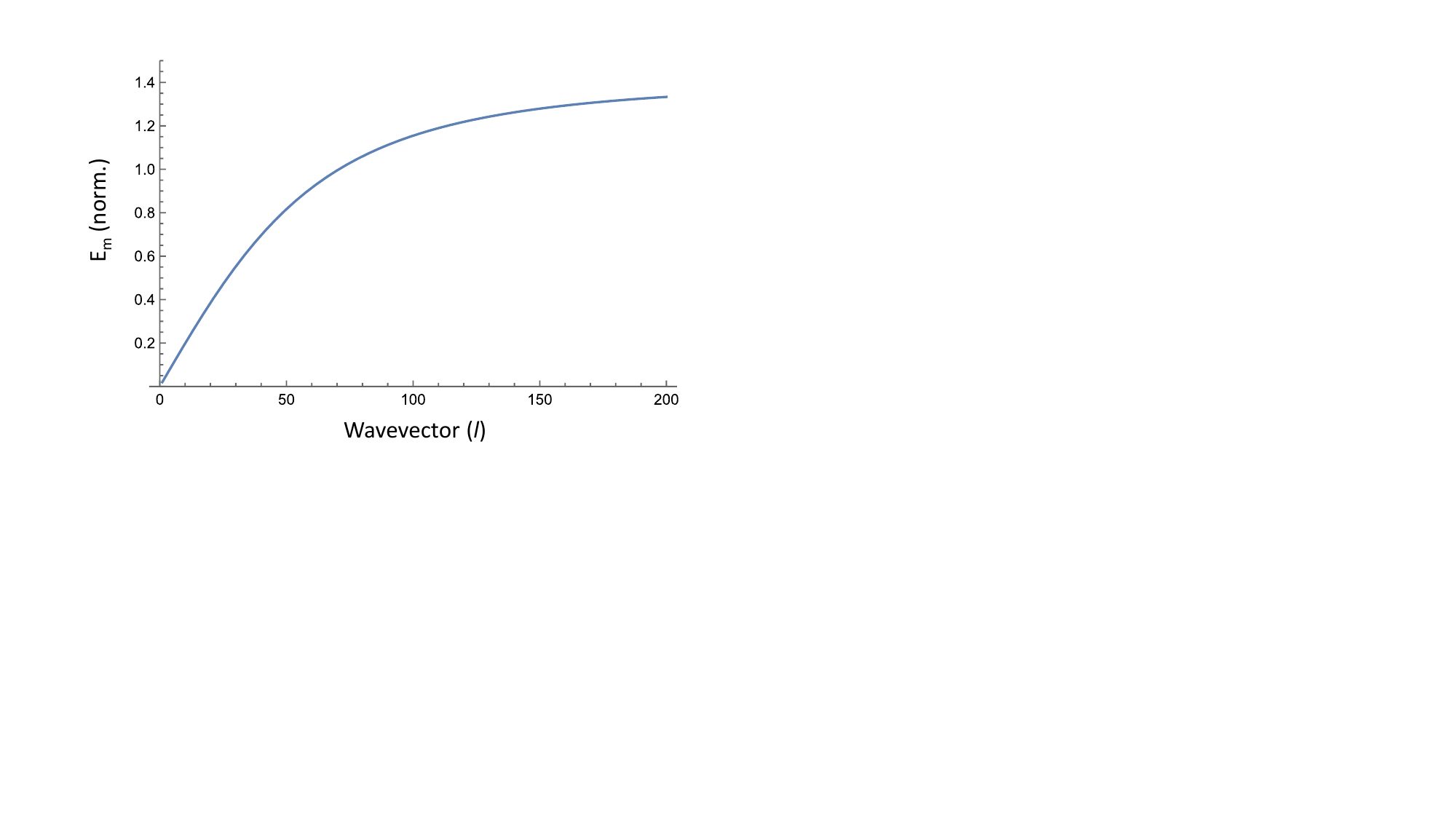}
    \vspace*{-7.2cm}			
			\caption{\small{ 
            TM$_{n,m}$ wave electric field quantum fluctuations $E_m$ normalized to $E_{zpf}$ as a function of wavevector number $l$, for branch $n=1,m=1$. We chose for the plot $w=d$ and $l=100\,d$. Asymptotes are $\sim 0.02$ at $l=1$ and $\sqrt{2}$ at $l\rightarrow + \infty$.}}
			\label{fig_2}
		\end{figure}

Eq. (\ref{bosons}) tells us that photons are {\it bosons}, as it should. Eq. (\ref{hbarval}) links the flux prefactors carrying the units $\phi_m, \phi_m'$ to the fundamental variable $\hbar$. It is nothing but our common definition of {\it zero-point fluctuations}, here in Volts.seconds.
It can be re-expressed in terms of the electric field amplitude $E_m$ (or the magnetic field $B_m=E_m/c$):
\begin{equation}
E_m = \sqrt{\frac{\hbar \, \omega}{2 C_P \, h_{e\!f\!f}^2}} , 
\end{equation}
and the equivalent formula with primed quantities. We give in Tab. \ref{tab_16} the calculated values for all configurations, using our top/bottom electrodes writing convention. The natural scale for electric field quantum fluctuations is thus $E_{zpf} \propto \sqrt{\omega}$; interestingly, it involves the full volume $d\,w\,L$ of the waveguide. For TE$_{n,m}$ and TM$_{n,m}$ modes with $n,m \in \N^*$, there is a prefactor which depends on the branch indexes $n,m$. 
But specifically for TM traveling signals, we also find out a $\propto \sqrt{\frac{\beta^2}{\beta^2+k_c^2}}$ dependence. This is quite peculiar: it means that as $|\beta| \rightarrow 0$, electric/magnetic field quantum fluctuations {\it tend to vanish} as $\propto |\beta|$. They recover the conventional $\propto \sqrt{\omega}$ dependence as $|\beta|\rightarrow \infty$. This is illustrated in Fig. \ref{fig_2}. \\

To our knowledge, this phenomenon has not been commented in the literature so far. We claim that a quantitative measurement of zero-point fluctuations as a function of wavevector, for a given TM$_{n,m}$ branch in a rectangular guide, could prove our result.

\section{Conclusion}

We presented the thorough quantum modeling of waveguides having the simplest Cartesian geometry, namely parallel plates and rectangular pipes. 
For TEM, TM$_n$ and TE$_n$ (equivalently TE$_{n,m=0}$ or TE$_{n=0,m}$) traveling modes, only one set of electrodes is involved in the description of the field. 
We introduce the notion of {\it virtual electrodes}, which applies to boundaries fulfilling the conditions corresponding to metallic plates, even in their absence. 
For TM$_{n,m}$ and TE$_{n,m}$ signals, two sets of equivalent electrodes are present (top/bottom and left/right), which can be equivalently referred to.

We give the analytic expressions for all fields, the real-space ones $\vec{E},\vec{B}$ but also the potentials $\vec{A}, V$.
This enables us to extend the generalized 
flux of M. Devoret $\dot{\varphi}=\Delta V$ \cite{devoret97} to all field configurations, beyond the usual TEM case. 
This is realized through a specific {\it gauge fixing} that we shall call the "phi-gauge", leaving only very little {\it gauge invariance} (at most 2 real coefficients).

All measurable physical quantities are recast in terms of the generalized 
flux $\hat{\varphi}$. It is interpreted as the {\it scalar field} which confines light in the waveguide, with quantum conjugate variable $\hat{Q}$, the {\it charge localized at the boundaries}, even when these are virtual. 
As such, {\it nothing is imposed} on the nature of these charges: these must simply correspond to particles that couple to the electromagnetic field. 

The low frequency cutoff $\omega_c = c \, k_c$ of non-TEM waves can be straightforwardly interpreted: it corresponds to an {\it  addendum energy} necessary for the addition of photons in the guide for TM waves, while it is linked to a {\it photon mass} in the TE configurations, arising from a {\it peripheral circulating} current around the guide. 
In the former case, the  $\hat{\varphi}$ propagation equation involves directly the phase velocity $v_\phi$, while in the latter it is the speed of light $c$ (as for TEM); but then the equation is a Klein-Gordon propagation expression, with a mass term verifying $m c^2 =\hbar \omega_c$.

The electric field quantum fluctuations are calculated for each type of propagating solutions. It turns out that 
for TM waves, a scaling $\propto \sqrt{\beta^2 /(\beta^2 + k_c^2)}$ appears, with $\beta$ the wavevector. 
It results in a {\it decrease} of the quantum fluctuations close to the cutoff (for $|\beta|\rightarrow 0$).
A measure of this effect would prove that the present approach is meaningful; it might also be used in order to transmit quantum information with especially low quantum noise modes. How to implement this in practice remains to be sorted out. \\

We believe that the presented calculations are both useful and enlightening, with many potential extensions. 
One can readily compute the total quantum field fluctuations $\sqrt{<\vec{E}^2>}$, presenting a typical scaling $\propto \omega_{max}/\sqrt{d\,w}$. It directly depends on the cross-section of the guide, and diverges as a high frequency cutoff $\omega_{max}$. This is not as problematic as for the free field quantum fluctuations (the famous zero-point energy divergence \cite{landau}), since the waveguide will stop being ideal at some finite frequency (so the cutoff has here a physical meaning). 
Besides, this calculation is directly linked to a fundamental phenomenon which it could help recast: electromagnetic quantum fluctuations are at the core of the (static) {\it Casimir effect} \cite{casimir1}, a force appearing between conductors at very small lengthscales. 
To be quantitative, the modeling must include the electrode's  electric properties \cite{casimir}, which is also an issue which could extend the present work.

The reasoning we developed here shall be adapted to cylindrical geometries. 
This is of particular practical importance, since many experiments rely on coaxial lines. 
The extension to free fields \cite{cohen,landau} (and especially Gaussian light \cite{menzel}) is also particularly relevant, but requires more thinking. 
Indeed, including the angular momentum of light, {\it both spin and orbital}, is a difficult but fundamental issue which is still a subject of research and debate today \cite{PRRspin}. 
Note also that since the theory is constructed in the first place from the Lorenz gauge, it is presumably compatible with a covariant treatment of the Hamiltonian \cite{cohen}; this deserves to be carefully analyzed.
Finally, a natural prolongation of the modeling would be to consider photons as {\it wavepackets} localized in space and time, 
so-called "quantum pulses"  \cite{photonpulse}.

\section*{Acknowledgements}

The theoretical developments presented here have been conducted within the framework of the European Microkelvin Platform (EMP) collaboration, 
visit: \underline{https://emplatform.eu/}.
It has emerged naturally from an
ERC CoG grant, ULT-NEMS No. 647917, which was dealing with microwave optomechanics and mechanical objects cooled to their quantum ground state of motion. 
Certainly many fruitful discussions with great colleagues have energized this thinking, but giving a fair list of names is quite impossible; we therefore simply wish to thank the community as a whole.


\newpage
\vspace*{2cm}

\begin{center} 
{\Large \bf Corrigendum: Waveguides in a quantum perspective \\
New J. of Phys. {\bf 28}, 059501 (2026) \\}
\end{center}
\vspace*{0.5cm}

\begin{center} 
  E. Collin$^{*}$, A. Delattre$^{*}$
\end{center}

\begin{center}
{(*) Univ. Grenoble Alpes, Institut N\'eel - CNRS UPR2940, 
25 rue des Martyrs, BP 166, 38042 Grenoble Cedex 9, France \\ }
\end{center}
\vspace*{1cm}

While working on follow-up articles, we realized that some expressions in the manuscript {\em waveguides in a quantum perspective} were slightly misleading. As well, some formulas could have been further simplified into more generic forms, helping in the understanding.

What is named "inverse inductance per unit surface" $L_d^{-1}$ is actually defined in Henry$^{-1}$, and should therefore appear in quotation marks. This is also true for the deduced "mode inductance" $L_H$, which is in fact in Henry$/$m$^2$. The mode inductance (in the accepted sense) is actually $L_H/\beta^2$, because of the derivative $\partial \varphi_{max}/\partial z$ that appears in the Hamiltonian expression, Eq. (112). 
As a matter of fact, what was meant by the original text is "the quantity related to an inductance", which would be a density or a characteristic of the mode. This abuse of language does not alter the conclusions, but could hinder the understanding of some readers.

Similarly, the velocity that appears in Tab. 16 should be understood as "the velocity entering the Klein-Gordon propagation equation of $\varphi$". This is actually stated in the conclusion, but is quite misleading: obviously, the phase velocity of {\em both} TM and TE waves is $v_\phi = c \, k/|\beta| \neq c$, and only for TEM do we have $v_\phi =c$.

Note also that the $\hat b, \hat b^\dagger$ parameters introduced are implicitly $t=0$ operators, in a Heisenberg picture [7]; obviously, their time-evolved counterparts are $\hat b \,e^{-i \omega t}$ and $\hat b^\dagger \,e^{+i \omega t}$ respectively, see Eqs. (122,123).
Furthermore, Eq. (64) and Eq. (73) while being correct, can be further simplified by rewriting them with the replacement:
\begin{eqnarray}
\frac{ 1 + k_{cx}^2/k_{cy}^2}{2} & \rightarrow & 1 , \\
\frac{ 1 + k_{cy}^2/k_{cx}^2}{2} & \rightarrow & 1 ,
\end{eqnarray}
in each of the integrals (over $x$ and $y$, respectively). This equivalency (when summing over each electrode's transverse direction) is in fact a symmetry property of the horizontal/vertical electrodes.
As a result, it is more convenient to define for the surface potential energies in the TM rectangular guide case, Eqs. (67,68):
\begin{eqnarray}
\Delta & = & \frac{1}{2} \, C_d \,  \left( \frac{k}{\beta}\right)^2 (c\, k_c )^2 \varphi^2 , \\
\Delta' & = & \frac{1}{2} \, C_d' \,  \left( \frac{k}{\beta}\right)^2 (c\, k_c )^2 \varphi'^2 .
\end{eqnarray}
As well, the charge/current energy density of the TE rectangular guide waves, Eqs. (76,77),  can be expressed in a more symmetric manner:
\begin{eqnarray}
H_d & = & \frac{1}{2} \, C_d^{-1} \, \sigma_t^2 +  \frac{1}{2} \, L_d \,\, j_{tz}^2 + \frac{1}{2} \, L_d  \left(1+ \frac{k_{cy}^2}{k_{cx}^2} \right)^{\!\!-1} \!\! j_{tx}^2  , \\
H_d' & = & \frac{1}{2} \, C_d'^{-1} \, \sigma_l^2  +  \frac{1}{2} \, L_d' \,\, j_{lz}^2 + \frac{1}{2} \, L_d'  \left(1+ \frac{k_{cx}^2}{k_{cy}^2} \right)^{\!\!-1} \!\! j_{ly}^2,
\end{eqnarray}
with $\, L_d  \left(1+ \frac{k_{cy}^2}{k_{cx}^2} \right)^{\!\!-1}$ and $ L_d'  \left(1+ \frac{k_{cx}^2}{k_{cy}^2} \right)^{\!\!-1}$ specific inductance densities associated to the peripheral currents. This makes the reading easier, and does not alter any of the conclusions.
\vspace*{8cm}


\end{document}